\documentclass[conference]{IEEEtran}
\usepackage{fancyhdr}
\IEEEoverridecommandlockouts
\usepackage{float} 
\usepackage{cite}
\usepackage{amsmath,amssymb,amsfonts}
\usepackage{algorithmic}
\usepackage{graphicx}
\usepackage{textcomp}
\usepackage{epsfig,endnotes}
\usepackage{booktabs}
\usepackage{fontawesome}
\usepackage{subcaption}
\usepackage{array}
\usepackage{multirow}
\usepackage{svg}
\usepackage{tikz}
\usepackage{amsmath}
\usepackage{listings, multicol}
\usepackage{filecontents}
\usepackage{xspace}
\usepackage{tabularray}
\usepackage{enumitem}
\usepackage{subcaption}
\usepackage{url}
\usepackage{stfloats}
\usepackage{hyperref}
\usepackage{soul}

\renewcommand\hl[1]{#1}

\makeatletter
\newcommand{\linebreakand}{%
    \end{@IEEEauthorhalign}
    \hfill\mbox{}\par
    \mbox{}\hfill\begin{@IEEEauthorhalign}
}
\makeatother

\lstdefinestyle{mystyle}{
    backgroundcolor=\color{backcolour},   
    commentstyle=\color{codegreen},
    keywordstyle=\color{magenta},
    numberstyle=\tiny\color{codegray},
    stringstyle=\color{codepurple},
    basicstyle=\ttfamily\footnotesize,
    breakatwhitespace=false,         
    breaklines=true,                 
    captionpos=b,                    
    keepspaces=true,                 
    numbers=left,                    
    numbersep=5pt,                  
    showspaces=false,                
    showstringspaces=false,
    showtabs=false,                  
    tabsize=2
}

\definecolor{codegreen}{rgb}{0,0.6,0}
\definecolor{codegray}{rgb}{0.5,0.5,0.5}
\definecolor{codepurple}{rgb}{0.58,0,0.82}
\definecolor{backcolour}{rgb}{0.95,0.95,0.92}
\definecolor{pinkcolour}{HTML}{de90c0}

\definecolor{redish}{RGB}{232,119,62}
\newcommand*\nicstep[1]{\tikz[baseline=(char.base)]{%
            \node[white,shape=circle,fill=redish,draw,inner sep=1pt] (char) {\color{white}\sffamily #1};}}

\definecolor{blueish}{RGB}{32,156,210}
\newcommand*\dpastep[1]{\tikz[baseline=(char.base)]{%
            \node[white,shape=circle,fill=blueish,draw,inner sep=1pt] (char) {\color{white}\sffamily #1};}}

\def\BibTeX{{\rm B\kern-.05em{\sc i\kern-.025em b}\kern-.08em
    T\kern-.1667em\lower.7ex\hbox{E}\kern-.125emX}}

\begin{document}

\title{Network-Offloaded Bandwidth-Optimal \\ Broadcast and Allgather for Distributed AI}

\author{\IEEEauthorblockN{Mikhail Khalilov}
\IEEEauthorblockA{\textit{Department of Computer Science} \\
\textit{ETH Zurich}\\
Zurich, Switzerland \\
mikhail.khalilov@inf.ethz.ch}
\and
\IEEEauthorblockN{Salvatore Di Girolamo}
\IEEEauthorblockA{
\textit{NVIDIA Corporation}\\
Zurich, Switzerland \\
sdigirolamo@nvidia.com}
\and
\IEEEauthorblockN{Marcin Chrapek}
\IEEEauthorblockA{\textit{Department of Computer Science} \\
\textit{ETH Zurich}\\
Zurich, Switzerland \\
marcin.chrapek@inf.ethz.ch}
\linebreakand
\IEEEauthorblockN{Rami Nudelman}
\IEEEauthorblockA{
\textit{NVIDIA Corporation}\\
\quad Santa Clara, United States\quad\\
ramin@nvidia.com\\}
\and 
\IEEEauthorblockN{\quad Gil Bloch}
\IEEEauthorblockA{
\textit{\quad NVIDIA Corporation}\\
\quad Yokne'am Illit, Israel \\
\quad gil@nvidia.com\\}
\and
\IEEEauthorblockN{Torsten Hoefler}
\IEEEauthorblockA{\textit{Department of Computer Science} \\
\textit{ETH Zurich}\\
Zurich, Switzerland \\
torsten.hoefler@inf.ethz.ch}
}

\maketitle
\pagenumbering{arabic} 
\pagestyle{plain}

\lhead{}
\rhead{}
\chead{}
\lfoot{\footnotesize{SC24, November 17-22, 2024, Atlanta, Georgia, USA
\newline 979-8-3503-5291-7/24/\$31.00 \copyright 2024 IEEE}} \rfoot{}
\cfoot{}
\renewcommand{\headrulewidth}{0pt} \renewcommand{\footrulewidth}{0pt}

\begin{abstract}

In the Fully Sharded Data Parallel (FSDP) training pipeline, collective operations can be interleaved to maximize the communication/computation overlap. In this scenario, outstanding operations such as Allgather and Reduce-Scatter can compete for the injection bandwidth and create pipeline bubbles. To address this problem, we propose a novel bandwidth-optimal Allgather collective algorithm that leverages hardware multicast. We use multicast to build a constant-time reliable Broadcast protocol, a building block for constructing an optimal Allgather schedule. Our Allgather algorithm achieves 2$\times$ traffic reduction on a 188-node testbed. To free the host side from running the protocol, we employ SmartNIC offloading. We extract the parallelism in our Allgather algorithm and map it to a SmartNIC specialized for hiding the cost of data movement. We show that our SmartNIC-offloaded collective progress engine can scale to the next generation of 1.6 Tbit/s links.

\end{abstract}


%-----------------------------------------------------------------------
\section{Introduction}
%-----------------------------------------------------------------------

Distributed AI training can be understood as a networking problem. For instance, FSDP, a recently proposed distributed training system, addresses model size limitations by sharding parameters, gradients, and optimizer states~\cite{zhao2023pytorch, rasley2020deepspeed, rajbhandari2020zero, dubey2024llama}. Data parallel workers, running on compute accelerators like GPUs, retrieve sharded weights using the Allgather operation and use the Reduce-Scatter operation to reduce and shard gradients~\cite{walker1996mpi}. In FSDP pipeline, Allgather and Reduce-Scatter operations on independent shards can be interleaved to maximize computation on GPUs with communication. This necessitates a robust underlying collective stack capable of efficiently progressing multiple in-flight collectives. Send bandwidth contention between operations can result in pipeline bubbles and decrease system efficiency.

Collective algorithms supported by state-of-the-art stacks rely on \textit{point-to-point} (P2P) communication between processes~\cite{nccl, oneccl, ucc, rccl, zhou2023accelerating}. While conventional P2P-based Allgather schemes (e.g., ring) achieve an optimal bound on schedule time~\cite{thakur2003mpichcolls, chan2007collective, nccl}, they are not optimal in terms of \textit{total data movement} across the network. For Allgather, P2P communication implies that in any algorithm schedule with $P$ participants, the same send buffer will be sent at least $P-1$ times~\cite{alexandrov1995loggp}. With respect to the FSDP algorithm, traffic reduction within the Allgather schedule can increase the throughput of other in-flight collectives, such as Reduce-Scatter.

\begin{figure}[]
    \centering
    \includegraphics[width=\columnwidth]{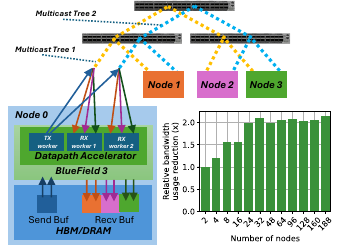}
    \caption{A simplified overview of the bandwidth-optimal Allgather algorithm represented as a composition of multicast-based Broadcasts. Multicast traffic processing is handled by the Datapath Accelerator. In the example above, the traffic is evenly distributed across two parallel multicast trees. We accommodate the discrepancy between data movement work on the send and receive paths by assigning one send and two receive path workers.}
    \label{fig:child_poster}
\end{figure}

To achieve this reduction, we propose a novel algorithm for Allgather, which has the per-process send bandwidth requirement scaling linearly with the send buffer size. We achieve this property by utilizing the hardware multicast feature. Specifically, we use hardware multicast to construct a constant-time Broadcast protocol. This protocol is a fundamental building block to express a bandwidth-optimal Allgather algorithm, which we realize as a composition of Broadcasts. With our algorithm, links in a Fat-Tree topology will transfer any byte of any participant's send buffer only once.

While the multicast-based Allgather scheme seems intuitive, the practical implementation of scalable multicast-based all-to-all-like communication is challenging and remains unsolved:
\begin{enumerate}[leftmargin=*]
\item The existing Remote Direct Memory Access (RDMA) interconnects support multicast offloading with datagram-based transports. Multicasting requires CPU involvement for each sent and received datagram. With a 200-400 Gbit/s link, a 4 KiB datagram size, and a single CPU core dedicated to the networking stack, the collective progress engine needs to sustain processing millions of datagrams per second simultaneously, progress other in-flight collective operations that require computation (e.g., Reduce-Scatter), and maintain the 99\% computation/communication overlap~\cite{belay2014ix, zhao2023pytorch, danalis2005transformations}.
\item The existing RDMA transports that support multicast are unreliable \cite{ibspec}. The added reliability protocol must minimally impact the rate of datagram processing.
\end{enumerate}

We address these challenges with a two-component protocol design: a highly parallel, fast-path layer that processes multicast datagram sends and receives (see Figure~\ref{fig:child_poster}) and a slow-path reliability layer based on the conventional ring algorithm and reliable one-sided RDMA operations. Since we design our protocol to be deployed with general-purpose lossless RDMA networks, the slow-path layer is triggered only in case of rare fabric drops. The fast-path layer leverages streaming processing in the data movement path of our protocol~\cite{hoefler2017spin}. We distribute the traffic across multiple multicast trees (i.e., streams), each carrying an independent part of the send buffer. The trees are processed in parallel by a multi-threaded progress engine.

To address deployment scenarios where the CPU cycle budget is scarce, we offload the progress engine to a SmartNIC. We utilize the Datapath Accelerator (DPA) \cite{chen2024demystifying-bf3dpa}, which is part of the ConnectX-7 NIC \cite{bf3}, as a SmartNIC offloading substrate. The key feature of DPA is its multi-core energy-efficient architecture with support for hardware multithreading. DPA consists of 16 RISC-V cores with 16 hardware threads per core. We efficiently leverage hardware multithreading to hide the latency of the fast-path layer, which mainly involves low-IPC data movement operations.

\textbf{Our main contributions are:}
\begin{enumerate}[leftmargin=*]
\item Analysis of data movement bottleneck in point-to-point-based collective stacks.
\item A novel Allgather algorithm based on hardware RDMA multicast and its open-source implementation.
\item An end-to-end prototype for a SmartNIC-offloaded collective progress engine tailored for Tbit/s link speeds.
\item System design principles for offloading networking stacks running on top of unreliable RDMA transports.
\end{enumerate}

%-----------------------------------------------------------------------
\section{Distributed training as a networking problem}\label{sec:background}
%-----------------------------------------------------------------------

We focus on optimizing the collective stack as one of the fundamental elements that impact the system's scalability.

\subsection{Injection bandwidth bottleneck}

With conventional Distributed Data Parallel training~\cite{li2020pytorch, ben2019demystifying, kurth2017deep}, the model size is limited by the size of GPU memory. The Fully Sharded Data Parallel (FSDP) approach applied to the model state (weights, gradients, and optimizer states) helps to overcome this constraint~\cite{zhao2023pytorch, dubey2024llama}. Data parallel workers invoke the Allgather operation during the forward and backward propagation to fetch sharded parameters and use the Reduce-Scatter operation after the backward pass to synchronize gradients.

The FSDP pipeline extensively overlaps the compute pipeline stages with collective communication between workers so that they are always saturated with training data. To achieve this, the pipeline leverages non-blocking collective semantics. Further, in the absence of data dependencies, Allgather and Reduce-Scatter operations can also progress simultaneously. Thus, having bandwidth-efficient algorithms for these collectives is a vital feature, as both operations can compete for the network injection bandwidth.

The existing production-grade collective backends utilize point-to-point (also known as unicast) primitives to express the Allgather algorithm. For example, the NCCL library uses the ring algorithm~\cite{nccl}. While point-to-point-based Allgather yields the lower bound on the time of the collective operation, the resulting communication schedule is not optimal regarding total data movement across the network. Let's assume there are $P$ participants in Allgather communication across a system with a Fat-Tree topology~\cite{al2008scalable}. Each participant posts a send buffer of size $N$ and a receive buffer of size $N \cdot P$ bytes. The linear Allgather algorithm requires $P - 1$ P2P connections per node and $N \cdot (P - 1)$ bytes to send per process (i.e., each participant has to send the same data to $(P-1)$ destinations)~\cite{alexandrov1995loggp}. Ring, tree, and recursive Allgather schemes help to reduce the number of connections per process to $1$ or $\log(N)$ \cite{thakur2003mpichcolls}. Yet, these algorithms do not improve data movement work on the send path: each participant needs to propagate the data to its neighbors/leaves.

\begin{figure}[ht!]
    \centering
    \includegraphics[width=\columnwidth]{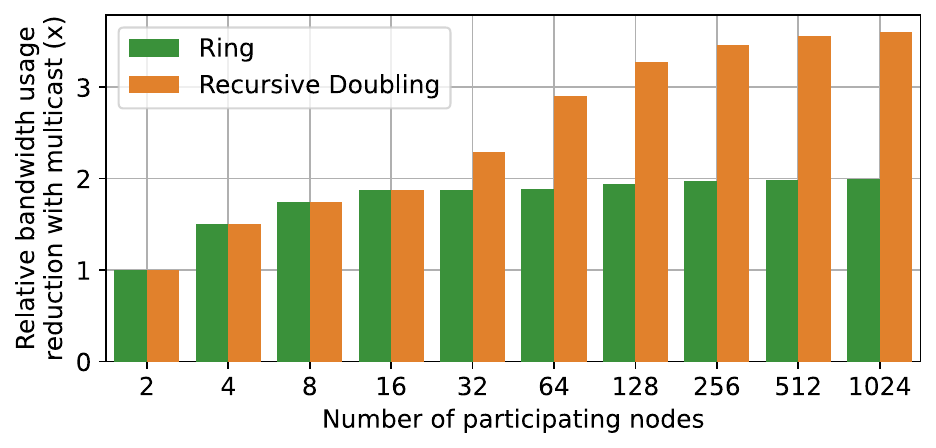}
    \caption{Theoretical cost model of bandwidth savings that can be achieved with multicast-based Allgather algorithm compared to classical point-to-point based approaches. The modeled system is a 1024-node cluster connected with a Fat-Tree topology using radix 32 switches.}
    \label{fig:theoretical_bw_savings}
\end{figure}

Multicast complements a unicast primitive to offload to the network the burden of distributing the same piece of data (e.g., send buffer) across multiple destinations. The switches propagate multicast traffic across the fabric endpoints that are attached to the same multicast \textit{group}. We can use multicast to build a reliable bandwidth-optimal constant-time (for a fixed buffer size) Broadcast algorithm. Further, we can compose Broadcasts from multiple senders in a bandwidth-optimal Allgather schedule. Figure~\ref{fig:theoretical_bw_savings} presents the theoretical traffic reduction achieved with such an algorithm. With a $2\times$ lower network bandwidth usage, a multicast-based algorithm allows us to provide more network bandwidth for other collectives running on the system.

\hl{For example}, in the next-generation FSDP deployments our multicast-based Allgather could be complemented with the in-network compute (INC) Reduce-Scatter algorithm (e.g., SHARP). With $P$ processes in the full-bandwidth fat-tree, the expected speedup $S$ with our approach when compared to the standard combination of the ring Allgather and Reduce-Scatter algorithms~\cite{gangidi2024rdma}, is $$S = 2 - 2/P,$$ meaning that \textit{at scale, the runtime of Reduce-Scatter and Allgather collectives concurrently initiated within the same set of nodes can be reduced by up to half}\footnote{See Appendix~\ref{sec:ag_cost_appendix} for analytical derivations.}.

\textit{\textbf{Insight 1:}} In general, any Allgather algorithm expressed with unicast primitive implies that at least one process needs to send $\Omega(N \cdot (P - 1))$ bytes. The multicast-based algorithm that we present in Sections~\ref{sec:broadcast} and~\ref{sec:allgather_offloading} ensures that \textit{the send buffer from any participant will be moved through any link in the network with Fat-Tree topology once}. We call collective algorithms that have this property \textbf{bandwidth-optimal}.

\begin{figure}[ht!]
    \centering
    \includegraphics[width=\columnwidth]{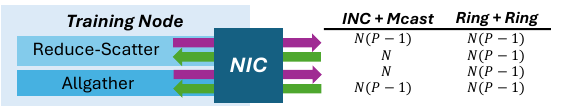}
    \caption{\hl{Data movement} at the training node boundary.}
    \label{fig:data_movement}
\end{figure}

\textit{\textbf{Insight 2:}} \hl{The bandwidth requirement of} conventional ring Allgather and Reduce-Scatter algorithms is equal on the send and receive paths (see Figure~\ref{fig:data_movement}). Thus, bandwidth paths of NIC are equally shared between these collectives. Instead, INC Reduce-Scatter is bounded by the NIC's send path bandwidth (to propagate send buffers to the network spine/core), while multicast-based Allgather has a bottleneck on the receive path. In other words, bandwidth-optimal Reduce-Scatter and Allgather algorithms don't share network bottlenecks.

\textit{\textbf{Challenge 1:}} Realizing bandwidth-optimal collective algorithms would require redesigning existing point-to-point-based protocol stack to leverage hardware multicast.

\subsection{RDMA and hardware multicast}

AI supercomputers widely leverage Remote Direct Memory Access (RDMA) interconnects for high-bandwidth data movement between memories of compute nodes~\cite{cscs_alps,kurth2017deep,meta_ai_center_wagner,de2020depth, li2019evaluating}. The InfiniBand specification~\cite{ibspec}, the most widely adopted RDMA implementation, offers three transport layer service models supported by the fabric endpoint called a Queue Pair (QP): Unreliable Datagram (UD), Unreliable Connection (UC), and Reliable Connection (RC). Multicast is standardized only for UD transport.

\begin{itemize}[leftmargin=*]
\item UD is the simplest transport in terms of NIC hardware implementation. It offers the unreliable delivery service of datagrams (in-order with drops) of Maximum Transmission Unit (MTU) size (up to 4 KiB). The UD queue pair has connection-less two-sided UDP-like semantics: datagrams can be sent to and received from any remote queue pair. UD can greatly improve scalability by maintaining a constant number of cached contexts in a small NIC SRAM cache. The corresponding UD QPs should be attached to a multicast group to send and receive multicast datagrams.
\item UC is another unreliable transport, but it supports arbitrary-length two-sided messaging and one-sided RDMA write messages. If one of the packets within the message is dropped, the entire message is also dropped. We also consider a possible extension of the next-generation RDMA hardware to support multicast with the UC transport, thus enabling multicast RDMA Writes.

\item RC alleviates the need for software reliability and re-transmission by offloading it to the hardware. This is a key feature to provide arbitrary-length one-sided operations used to implement the zero-copy rendezvous protocol, a fundamental building block for point-to-point-based collectives. However, since the RC transport requires a per-connection reliability state, it doesn't support datagram multicasting.
\end{itemize}

\begin{figure}[ht!]
    \centering
    \includegraphics[width=\columnwidth]{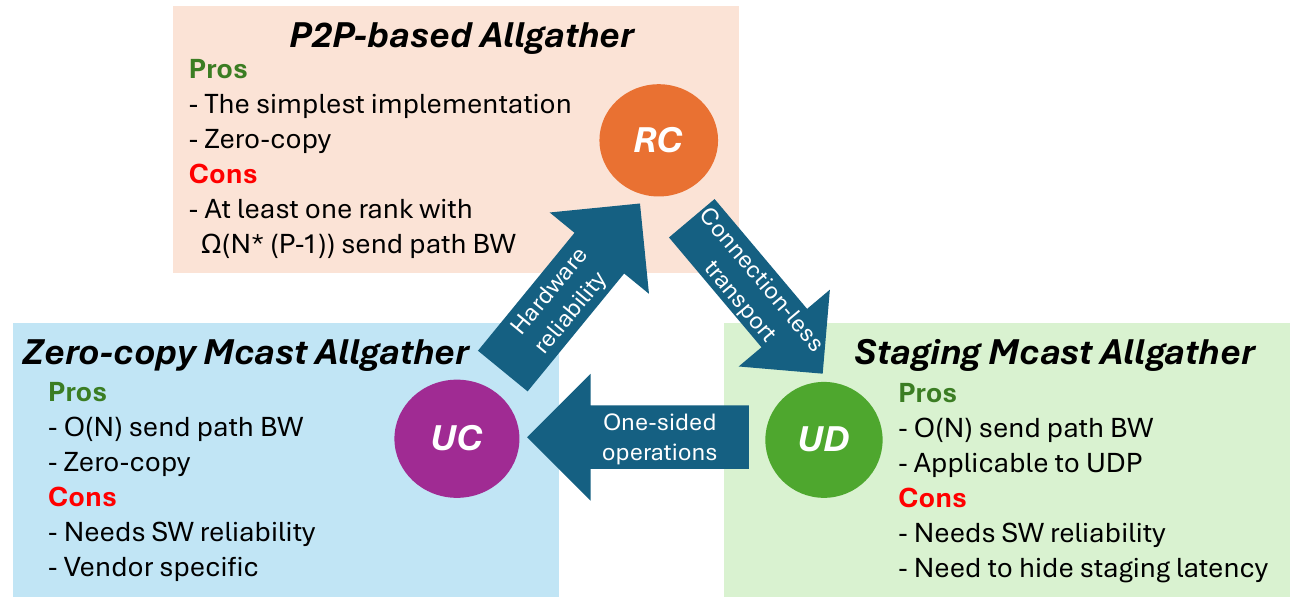}
    \caption{Trade-offs exist between the InfiniBand (IB) Verbs transport layer semantics and Allgather algorithm design. We present a practical multicast-based solution for Unreliable Datagram (UD) and Unreliable Connected (UC) transports.}
    \label{fig:transport_vs_ag_algo}
\end{figure}

\begin{figure}[ht!]
    \centering
    \includegraphics[width=\columnwidth]{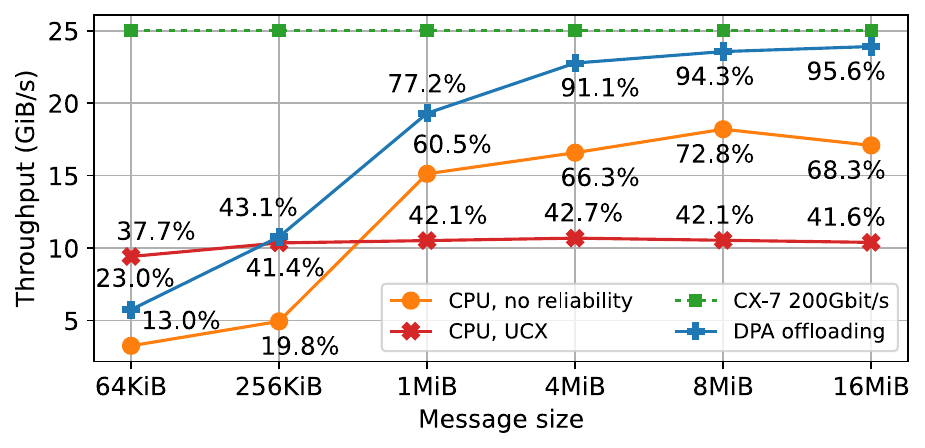}
    \caption{A single-threaded datagram-based datapath running on a server-grade CPU is unable to sustain the 200 Gbit/s link bandwidth, while the datapath offloaded to the single multi-threaded DPA core scales to the peak throughput.}
    \label{fig:cpu_bottleneck}
\end{figure}

We highlight the trade-offs between the QP service model and the resulting Allgather algorithm in Figure~\ref{fig:transport_vs_ag_algo}. For UD transport, the multicast feature requires software involvement on the per-datagram granularity to perform buffer segmentation and reassembly on the stack's sender and receiver sides. Multicast-enabled transports are also unreliable, and datagrams can be dropped. Therefore, the progress path must also support software reliability. In Figure~\ref{fig:cpu_bottleneck}, we assess the performance of the UD-based segmentation-and-reassembly and reliability protocols implemented in the production-grade point-to-point RDMA middleware (UCX). The experiment was conducted on a 2-node system equipped with server-grade 2.6 GHz AMD Epyc CPUs and 200 Gbit/s NVIDIA ConnectX-7 NICs (see Section~\ref{sec:evaluation} for more details). We observe that a single CPU core cannot reach the full link bandwidth. To cut out the software overheads of the reliability protocol, we also implement a custom progress engine where MTU-sized chunks are delivered through the RC transport. The single-threaded stack struggles to reach 200Gbit/s even without running the software reliability layer.

\textit{\textbf{Insight 3:}} With a lossless\footnote{Previous works \cite{hoefler2023data, dang2018lrum} report bit error rates for Ethernet and InfiniBand links to be on the order of $10^{-12}$ and $10^{-15}$, correspondingly.} link layer (e.g., InfiniBand or RoCEv2) and enough processing capacity on the receive side (to avoid Receiver-not-Ready drops), most of the time a TCP-like stack will be redundant. In Sections ~\ref{sec:broadcast} and ~\ref{sec:allgather_offloading}, we exploit traffic parallelism in the data flow of Broadcast and Allgather to realize the collective progress engine that can scale the datagram processing across worker threads.

\textit{\textbf{Challenge 2:}} The multi-threaded datagram processing will impose a significant CPU cycle footprint. These overheads must be avoided in application deployment scenarios where the computation is done on the CPU cores, e.g., distributed file systems~\cite{di2022building,kim2021linefs}.

\subsection{Datapath Accelerator}

We solve the CPU bottleneck by offloading the collective progress engine to the SmartNIC. We use Datapath Accelerator (DPA) as a SmartNIC offloading substrate. DPA is available in the latest generation of NVIDIA BlueField DPU, SuperNIC and ConnectX products. We describe the key DPA capabilities with respect to collective offloading.

The current DPA generation consists of 16 programmable energy-efficient RISC-V cores with 1.5 MB of last-level cache (LLC). The cores are clocked at 1.8 GHz and tailored to hide the cost of low instruction per-cycle (IPC) data movement code through hardware multi-threading. Each DPA core supports 16 hardware threads, resulting in 256 hardware execution contexts.

The DPA cores are directly interfaced with the NIC DMA engine and can be programmed using the DOCA FlexIO C API from the application user space. Using the DOCA FlexIO low-level API, the user defines a C kernel to be executed upon the completion queue events generated. Within the DPA kernel, the user can initiate data transfers using RDMA and Atomic operations. Any of the remote and local memory regions (with respect to the server where the NIC is installed) can be the targets of these operations. The user associates completion queues with the kernel so that the first completion event (e.g., send or receive operation completion) results in the execution (activation) of the event handler kernel. The hardware thread executes the activated kernel.

\textit{\textbf{Insight 3:}} Most of the cycles spent in the buffer segmentation and reassembly path of the collective progress engine correspond to the posting of the RDMA operations (stores) and polling for their completions (loads). In Section~\ref{sec:evaluation}, we show how the multi-threaded DPA architecture allows us to hide data movement latency in our collective progress engine and scale it to the current and next-generation link bandwidths.

\section{Constant-time reliable Broadcast protocol}\label{sec:broadcast}

In this section, we present the design of our protocol for reliable constant-time broadcasting on top of unreliable hardware multicast. The constant-time broadcasting protocol allows us to build the bandwidth-optimal Allgather schedule.

The first component of our protocol is the multicast fast path, which performs the segmentation and reassembly of the user buffer as fast as possible. In the fast path, the throughput is limited only by the rate at which the RDMA operations can be posted by the CPU or the SmartNIC thread and processed by the NIC DMA engine. Missed chunks are fetched using the reliable path. The rationale behind this two-component design is that in a lossless fabric drops are rare, and data will be delivered to the receivers through the fast path.

We describe the fast-path protocol in the context of UD transport. In the multicast path all the send and receive processes are attached to the same multicast group.

\subsection{Broadcast root: sender datapath}

The root process performs the fragmentation of the send buffer. It chunks up the user send buffer into MTU-sized datagrams and posts RDMA multicast send operations. The buffer fragmentation is zero-copy, so no intermediate copies are involved in the send path. Each buffer chunk (datagram payload) is associated with a packet sequence number (PSN) that enumerates the chunk within the send buffer. PSN is written into the 32-bit immediate data field of the posted RDMA send request, so it can be delivered to the receive side in the packet header.

In the absence of packet drops and assuming that all leaf participants pre-post receive buffers in advance (see the receiver-not-ready discussion in the next subsection), the datagrams posted by the root will be delivered to all leafs without the need to involve the reliability layer.

\subsection{Broadcast leaf: receiver datapath}

\begin{figure}[]
    \centering
    \includegraphics[width=\columnwidth]{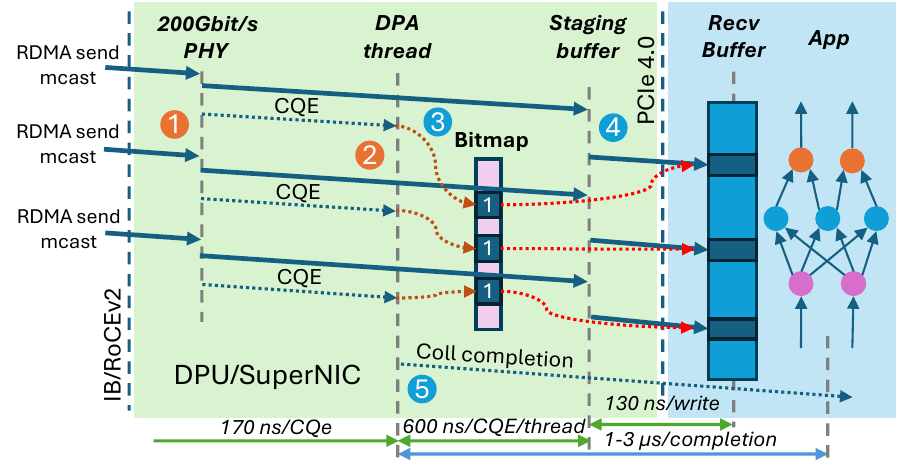}
    \caption{Break-down of receive datapath pipeline offloaded to the SmartNIC. We scale the number of receive workers to hide latency of single DPA thread. Orange and red arrows represent the relation between PSN stored in CQE immediate data, bitmap offset for the received chunk.}
    \label{fig:ud_pipeline}
\end{figure}

The leaf processes perform the buffer re-assembly. The corresponding datapath worker thread (e.g., running on the SmartNIC as shown in Figure~\ref{fig:ud_pipeline}) polls the network completion queue. During \nicstep{1}, each chunk is stored in the packet buffer resulting in a receive completion (CQE) generated by the NIC DMA engine in \nicstep{2}. Upon the completion being polled in \dpastep{3} by the worker thread, the corresponding chunk is marked as received. \dpastep{4} constitutes a memory copy of this chunk from the staging area to the user buffer. After all buffer chunks are received, the completion is generated, and the receive buffer can be released to the user in \dpastep{5}.

\textbf{\textit{Out-of-order traffic and packet drops:}} While the standardized UD semantics is in order, we expect that with a wider adoption of adaptive routing in the next generation networks, datagrams can be delivered to the receive side out-of-order~\cite{ibspec, hoefler2023data, de2020depth}. Thus, the user's receive buffer cannot be used to post the network receive requests and employ a zero-copy approach on the receive side without keeping track of the received packet order~\cite{koop2007zero}. For example, if the $i$'th chunk was dropped or re-ordered by the fabric, the $i+1$'th chunk will be matched to the $i$'th receive request. Assuming that we posted the user's receive buffer to the network, the $i+1$'th chunk will be stored at the $i$'th location in the user's receive buffer, resulting in its corruption.

Existing point-to-point protocols address this problem with the go-back-N or selective zero-copy fetches~\cite{koop2007zero, chu2016mcastreliability}. After detecting the out-of-order packet, the sender side is requested to re-transmit the buffer. While this approach allows support for zero-copy rendezvous over UD channel, the go-back-N scheme with multicast will require the delivery of go-back negative-acknowledgment (NACK) to the multicast root possibly from multiple sources, thus resulting in N-to-1 incast. The go-back-N scheme will also increase the software complexity of the receive datapath (e.g., the receiver state machine will include acknowledgement (ACK) and re-transmission recovery logic) and will not apply to out-of-order datagram delivery.

\textbf{\textit{Receive-side staging:}} We propose the staging-based solution, i.e., \dpastep{4} in Figure~\ref{fig:ud_pipeline}. First, each chunk is received into the staging memory area. The staging area is organized as a ring buffer. Each received chunk is copied from the staging buffer into the user's receive buffer. We rely on the DMA engine supporting non-blocking queuing to efficiently hide the latency introduced by additional memory copy in the UD datapath (e.g., $1-3 \mu$s PCIe latency in Step 4 in Figure~\ref{fig:ud_pipeline}), so that packet receives from the network \textit{to} the staging buffer can be overlapped with DMA copy \textit{from} the staging towards the user buffer~\cite{docadma,agarwal2022understanding}. Out-of-order delivery is also supported by design as PSN written in the CQE allows determining the offset of the chunk in the destination buffer.

\subsection{Slow-path: synchronization and reliability layer}

While we design our protocol to be deployed with the lossless link layer, in practice, the receive side can observe packet drops in two scenarios:
\begin{enumerate}[leftmargin=*]
\item \textit{Receiver-not-Ready (RNR) drops} occur when, by the time of multicast packet arrival to the destination NIC, the corresponding receive request was not posted yet. We avoid this kind of drops using two mechanisms. We pre-post the network receive queue with receive requests in receive buffer and then perform the barrier synchronization before the root starts broadcasting. To keep up with the re-posing of receive requests we scale the number of datapath workers on the receive side (see Section~\ref{sec:receive_side_scaling}).
\item \textit{Fabric drops} might occur because of packet corruption in the link layer. To recover in this scenario, each Broadcast leaf maintains the receive buffer bitmap and performs zero-copy fetch of missed chunks during the recovery phase.
\end{enumerate}

\begin{figure}[]
    \centering
    \includegraphics[width=\columnwidth]{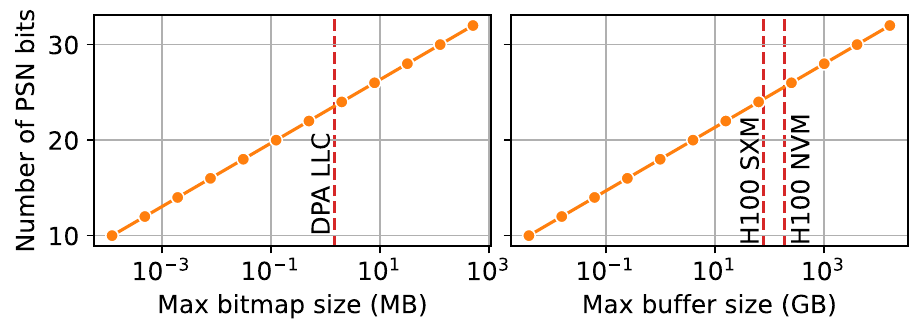}
    \caption{Maximum Allgather bitmap and receive buffer sizes are modeled as functions of PSN bits allocated in IB Verbs 32-bit CQE immediate value. We show device memory sizes for the current generation of NVIDIA GPUs and DPA.}
    \label{fig:psn_size}
\end{figure}

\begin{figure*}[h!]
\includegraphics[width=\textwidth]{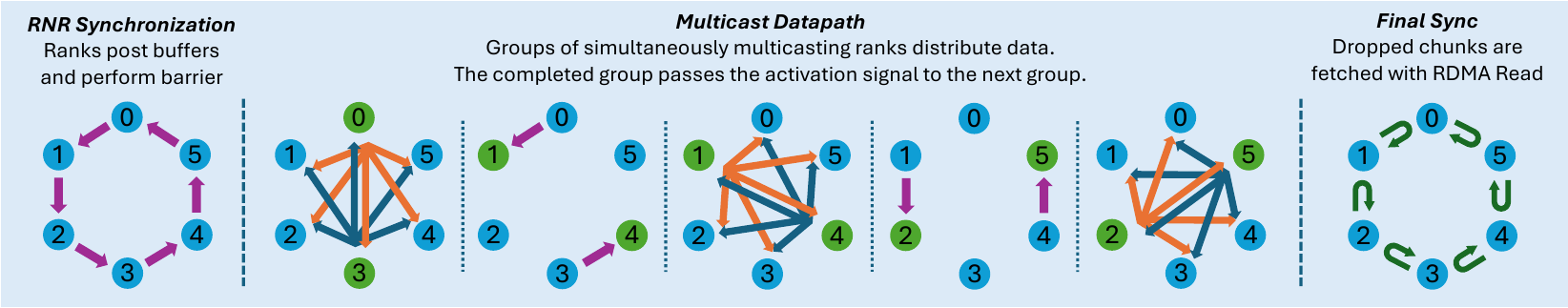}
\caption{Overview of the multicast-based Allgather algorithm schedule across six processes and two actively multicasting roots.}
\label{fig:allgather_schedule}

\end{figure*}

\begin{figure}[h!]
    \centering
    \includegraphics[width=\columnwidth]{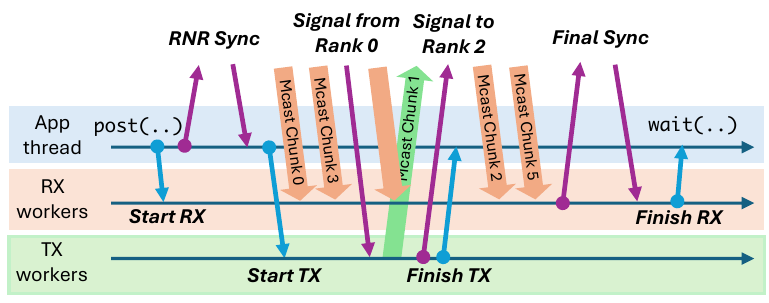}
    \caption{The Allgather execution flow is shown from the perspective of Process 1 in Figure~\ref{fig:allgather_schedule}. Violet arrows correspond to the protocol control path communication over the network. Turquoise arrows correspond to the atomic communication between the application and worker threads.}
    \label{fig:allgather_workers}
\end{figure}

\textbf{\textit{Bitmap:}} In our design, the receive datapath keeps track of each received chunk in a bitmap. The bitmap location is determined by the PSN stored within the receive completion immediate data. We choose the bitmap as a reliability data structure because it allows us to store information about drops in a compact way with minimal overhead on the receive datapath throughput. In Figure~\ref{fig:psn_size}, we analyze how many bits in the completion immediate data must be allocated to address the system memory while ensuring that the bitmap fits in the LLC of the SmartNIC that we use for evaluation\footnote{The remaining immediate value bits can be used to store implementation-specific information, such as the collective ID.}.

\textbf{\textit{Cutoff timer:}} When the leafs enter the multicast phase, they start polling receive completions. To account for extreme scenarios where some of the packets were dropped, the receive thread sets the timeout equal to $(N / B_{link} + \alpha)$ seconds, where $N$ is the receive buffer size and $B_{link}$ is the link bandwidth. The parameter $\alpha$ is adjusted to account for RNR synchronization time and networking noise.

\textbf{\textit{Fetch layer:}} We use RC transport to create a fetch ring across all processes. After the completion of the multicasting phase, leafs scan the bitmap. Leafs that have experienced fabric drops perform a customized version of the zero-copy rendezvous protocol with their neighbors in the virtual ring. For example, the process that misses a chunk will send a buffer fetch request to its left neighbor in the ring. If the process's left neighbor has all the buffer chunks, it will send the ACK packet back. After receiving the ACK, the process selectively fetches missing chunks utilizing RDMA Read primitive. If the process's left neighbor also experienced drops, it will recursively apply the fetch scheme until the leaf (Broadcast root in the worst case) that has all the chunks is found.

Our reliability scheme features several advantages. We avoid incasting the Broadcast root with ACK and fetch requests. We can easily generalize the scheme for multicast-based Allgather: in the worst-case scenario, it results in the ring Allgather that yields the optimal bound on the receive-side bandwidth. In the absence of packet drops, the processes need to perform only the final handshake in the reliable ring.

\textbf{\textit{Final handshake:}} When a leaf receives all chunks, it sends the final packet to its neighbor in a ring. Sending a final packet to the left process and receiving a final packet from the right process constitutes the final handshake. After the handshake, the receive buffer can be released to the application.

\subsection{Memory footprint analysis}

\paragraph{Connection contexts} In the fast path, a single multicast UD QP sends and receives data from all remote sources. The synchronization path is based on the ring algorithm and requires 2 RC QPs per leaf.
\paragraph{Staging area} The size of the staging area is bounded by the size of the QP receive queue and the rate at which the worker thread can process packets. For the BlueField 3 SmartNIC used in our experiments, the maximum receive queue size is 8192. With a 4 KiB MTU, the maximum staging area size is 32 MiB. In our experiments with DPA we observe that a practical size to sustain the 200 Gbit/s link is 4 MiB. We allocate this memory within the 16 GB DDR5 memory of the BlueField SmartNIC.
\paragraph{Bitmap} The bitmap is the only protocol state that grows linearly with the receive buffer size. As shown in Figure~\ref{fig:psn_size}, the bitmap size that fits in the DPA LLC (1.5 MB) will allow addressing the Allgather receive buffer of approximately 50 GB.
\paragraph{Multiple communicators} Assuming 64 KiB bitmap (i.e., up to 16 GB Allgather receive buffer) and 16 KiB per-communicator context size, more that 16 communicators will fit in the DPA LLC.

%-----------------------------------------------------------------------
\section{Allgather as a composition of Broadcasts}\label{sec:allgather_offloading}
%-----------------------------------------------------------------------

The key to the scalability of our system is its modular design. We use the Broadcast protocol described in the previous section as a building block to assemble the bandwidth-optimal Allgather algorithm. Intuitively, the Allgather schedule constitutes a round-robin scheduling of broadcasting processes across all collective participants.

Our Allgather algorithm must scale across the network resources (e.g., NIC RDMA engines and switches) and the protocol processing resources attached to the network endpoints (e.g., SmartNIC or CPU cores). We achieve this by hierarchically extracting the parallelism in our protocol:
\begin{enumerate}[leftmargin=*]
    \item \textit{Multicast parallelism:} We adjust the number of concurrently broadcasting processes that execute the protocol described in the previous section. Figure~\ref{fig:allgather_schedule} shows an example of Allgather algorithm execution.
    \item \textit{Flow direction parallelism:} With multiple broadcasting roots, the roots process traffic independently within different worker paths, as illustrated in Figure~\ref{fig:allgather_workers}.
    \item \textit{Packet parallelism:} We replicate the multicast groups to evenly distribute the packet processing across threads, with each worker thread processing a smaller portion of the broadcasted buffer.
\end{enumerate}

\subsection{Multicast parallelism: distributed scheduler}

When all Allgather participants start multicasting at the same time, this will result in heavy incast network congestion towards multicast group endpoints. Additionally, the switch implementation of multicast might require several multicast sub-trees to be active at the same time to keep the egress ports saturated with traffic. Thus, we need a mechanism to control the aggregate multicast traffic volume. While being lightweight, this mechanism must precisely manage the volume of traffic traversing the fabric at any given point in time.

\textit{\textbf{Broadcast sequencer:}} To solve this problem, we split the virtual ring of Allgather participants into $M$ parallel \textit{Broadcast chains} (see Appendix~\ref{sec:scheduler_appendix} for formal definition). The processes within chain will multicast the data one-by-one. We execute this scheduling within all chains in parallel. We use the synchronization layer to propagate the activation signal within a chain. For example, once a process finishes multicasting, it sends the activation signal to its neighbor in the chain. We can choose the chain split based on the network topology. For example, we can map chains to the server racks to limit the outbound multicast traffic per each rack.

\subsection{Flow direction parallelism: splitting send and receive paths}

We split the send and receive datapaths between different worker threads. With a multi-threaded design, we alleviate the need to balance the core cycle budget between traffic injection and reception. This is critical for the Allgather schedule stages where the processes act as Broadcast roots and leafs simultaneously.

We illustrate the synchronization flow in Figure~\ref{fig:allgather_workers}. We utilize atomic primitives for the control signaling between the application thread and workers to ensure the portability of the protocol across different SmartNIC and CPU architectures.

\textit{\textbf{Request submission:}} The operation is initiated when the application thread posts a new task. The task is added to the queue that is shared between the worker and application threads. After that, the receive path is signaled to start the cutoff timer and poll the network queue for the new datagrams. The application thread performs RNR synchronization with other processes and signals to the send path.

\textit{\textbf{Request progress:}} The sender worker executes the multicast scheduler. When it's turn in the schedule arrives, it starts to inject multicast packets. Once the send worker finishes multicasting, it signals to the application thread. On the receive side, once all the packets are successfully received by the receive worker, the final handshake is performed. In case the cutoff timer expires, the receive worker also executes the reliability recovery before the final handshake.

\subsection{Packet parallelism: distributing buffer across workers}\label{sec:receive_side_scaling}

The advantage of a multicast-based Allgather is that as the number of participants increases, the per-process send bandwidth requirement remains constant and equals the size of the send buffer. However, the receive bandwidth requirement scales with the number of participants. For example, suppose there are 16 processes, each sending an 8 MiB buffer. The receive path will need to handle a total of 120 MiB, resulting in 15$\times$ more work. Therefore, it's crucial to focus on optimizing the scalability of the Allgather receive side.

\textit{\textbf{Multicast subgroups:}} We evenly distribute the traffic across multiple multicast groups, referred to as \textit{subgroups}. Each worker thread polls the completion queue associated with one or more multicast subgroups. Mapping contiguous send buffer blocks to subgroup QPs allows us to maintain a thread-local bitmap on the receive side and limit inter-thread synchronization to thread activation and tear-down.

With this design, we can scale the number of workers on the send side independently from the receive side. For example, consider a scenario with 16 participating processes, 4 multicast subgroups, and an 8 MiB send buffer. On the send path, each participant serves contiguous 2 MiB buffer blocks across 4 QPs. On the receive path, each QP will handle 30 MiB. To address this discrepancy between sender and receiver, we can allocate 1 send worker serving all 4 send QPs, and 4 receive workers mapped one-by-one to the QPs.

%-----------------------------------------------------------------------
\section{Implementation}
%-----------------------------------------------------------------------

The goal of our implementation is to support two types of systems: current-generation machines leveraging standard CPU-driven RDMA offloading, and next-generation deployments based on programmable SmartNICs. We open-source the code for the community:

\centerline{\href{https://github.com/spcl/muliticast-based-allgather/}{https://github.com/spcl/muliticast-based-allgather/}}

\subsection{CPU-driven implementation in the UCC library}\label{sec:ucc_implementation}

We implement our multicast protocol as a backend for the open-source Unified Collective Communication (UCC) library. The UCC library is natively supported by OpenMPI and PyTorch runtimes. Our multicast-based UCC backend comprises $\approx$3500 lines of C code. Our backend supports two collective operations: Broadcast and Allgather. The modular architecture of our multicast-based Broadcast protocol allows us to use the same data and reliability path kernels for both collectives. The only difference is around 20 lines of code related to the Allgather multicasting scheduler.

We describe the key optimization techniques that helped us achieve a scalable implementation of our protocol.

\textit{\textbf{Initialization phase:}} We use C11 atomics to implement a lock-free task queue and non-blocking signaling between the main application thread and workers. To optimize the collective initialization phase, we cache the user buffer memory registrations, use a memory pool for the work requests, and also employ the recursive-doubling barrier in the RNR synchronization step.

\textit{\textbf{Progress engine:}} We employ RDMA-specific optimization techniques to reduce overheads on the send and receive worker datapaths~\cite{kalia2016rdmadesign, zambre2019breaking}. Send workers batch multicast requests to reduce the number of NIC DMA engine doorbell updates. Only the completion of the last send request in the batch is reported through the completion queue, reducing the number of reported completions. On the receive side, we pre-post the network receive queue with receives in the staging buffer and cache all staging receive work requests for fast re-posting.

\subsection{Offloading the protocol datapath to DPA}

We offload the collective progress engine using NVIDIA DPA to address scenarios where host CPU cycles are scarce and must be dedicated to the application (e.g., compute stages of the training pipeline run on the CPU).

We focus on the receive-side offloading because this is the main bottleneck on the data flow path in Allgather: the receive side needs to process data from all senders, keep track of reliability state, and copy chunks from the staging area in the BlueField memory to the user buffer. The corresponding instruction path consists predominantly of data movement tasks, such as reading completions, re-posting RDMA receives, updates to the bitmap, etc. This makes DPA the perfect choice for our goals, as we can hide the latency of the low-IPC code by running parallel workers on energy-efficient multi-threaded cores (see Figure~\ref{fig:ud_pipeline}).

We utilize the event-driven programming model supported by the DOCA FlexIO API~\cite{docaflexio}. As the semantics of interaction with the NIC RDMA engine in the DOCA FlexIO API is similar to the host-side Verbs API, the receive-side optimizations that we apply in our UCC backend directly translate to the DPA kernel code. The DPA-offloaded progress engine works as follows:
\begin{enumerate}[leftmargin=*]
    \item The DPA thread context is initialized after the host-side application registers memory using the standard IB Verbs API and copies the buffer metadata (e.g., pointers and memory keys) to the DPA thread memory.
    \item The completion of the first chunk reception, which corresponds to the pre-posted RDMA receive, activates the hardware DPA thread that starts executing the receive datapath kernel.
    \item The kernel polls the completions for subsequent packets. For each received chunk, the kernel sets the bit in the bitmap. The kernel issues an RDMA write operation through the loopback queue to copy the chunk staging area to the user receive buffer.
    \item Once all chunks are received, the last worker notifies the application thread about the completion of the operation by setting a flag in the host memory. After that, the receive buffer can be consumed by the application.
\end{enumerate}

To further minimize the software involvement we prototype the second version of receive datapath that is based on UC multicast (we present the kernel code in Appendix~\ref{sec:dpa_appendix}), a possible extension for the next-generation RDMA networks. As the UC transport supports arbitrary-length RDMA write messages, the staging on the receive side, which is necessary for UD, becomes redundant.

\subsection{Support for multiple communicators}

\hl{In our setup}, each new communicator is mapped to a set of threads. A single thread serves a group of parallel multicast trees, with each tree associated with a bitmap. Our UCC prototype implements a round-robin mapping of communicator progress threads to the cores. With enough communicators, the progress threads will eventually oversubscribe CPU cores, leading to a noticeable slowdown due to context switching. We anticipate that context switching overheads can be avoided by supporting software traffic arbitration. In this approach, each progress thread subscribes to multiple communicator contexts and serves traffic from them on a per-datagram basis.

\hl{The same approach} can be applied to DPA offloading. In Section~\ref{sec:dpa_perf}, we demonstrate that 16 DPA threads (out of 256) per communicator are sufficient to sustain a datagram arrival rate at 200 Gbit/s, indicating that DPA has enough compute capacity to also sustain software arbitration.

\section{Evaluation}\label{sec:evaluation}

We study the following questions:
\begin{enumerate}[leftmargin=*]
\item Does our multicast-based algorithm provide network bandwidth usage savings and throughput improvements compared to classical point-to-point-based collectives?
\item What is the raw performance of the SmartNIC-offloaded Allgather datapath?
\item Does SmartNIC-based offloading help overcome the CPU bottleneck in the protocol progress path?
\end{enumerate}

\subsection{Evaluated systems}

We follow scientific benchmarking guidelines when reporting the experimental results~\cite{hoefler2015scientific}.

\textit{\textbf{UCC testbed:}} We evaluate our UCC multicast backend on a 188-node testbed based on a fat-tree topology composed of 18 Mellanox SX6036 switches. The hosts are equipped with ConnectX-3 56 Gbit/s NICs, 10-core 2.20 GHz Intel Xeon CPU E5-2660v2, 32 GB of DDR3 DRAM, and run CentOS 7.3. We use OpenMPI v5.0.2 and compile it with UCX v1.16 and UCC v1.3.x collective backend support. We use OSU benchmarks v7.3 for benchmarking. We patch the benchmark to log the per-iteration times of the collectives across all MPI ranks. We compile the stack using GCC v11.4.0.

\textit{\textbf{DPA testbed:}} For experiments with SmartNIC acceleration, we use two servers connected back-to-back with the latest generation of BlueField 3 DPU. Our experiments use one of two 200 Gbit/s ports, as the server motherboard supports only PCIe 4.0. We use the 16-core DPA engine integrated with the BF-3 RDMA engine. DPA is interfaced with 16 GB of DDR5 DRAM managed by the BlueField ARM subsystem. Each DPA core supports 16 hardware threads. The host is provisioned with a 24-core 2.6 GHz AMD Epyc 7413 CPU, 256 GB DDR4 RAM, and runs Ubuntu 22.04. We use the compilation toolchain supplied with DOCA SDK v2.2.0.

\hl{With these testbeds we address address different research goals}. With the \textit{UCC testbed}, we assess the scalability of our multicast-based algorithms. With the \textit{DPA testbed}, we show that the receive datapath, a fundamental bottleneck for our Allgather algorithm, can be efficiently parallelized and offloaded to the SmartNIC.

\subsection{Performance of multicast-based Algorithms}

\begin{figure}[]
   \centering
    \includegraphics[width=\columnwidth]{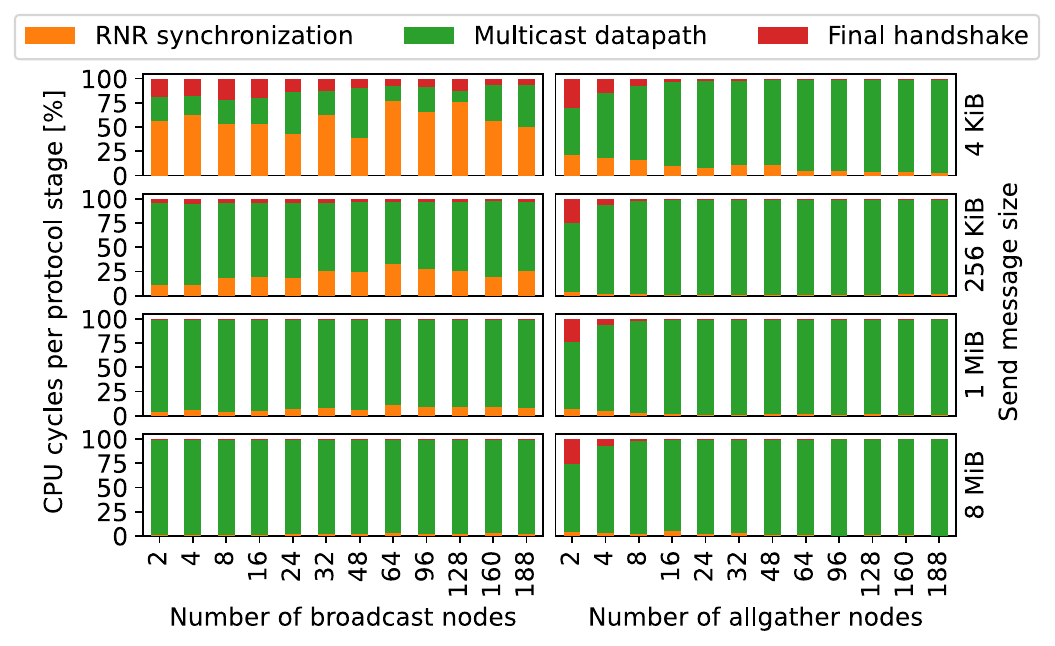}
    \caption{Protocol critical path break-down.}
    \label{fig:critical_path_breakdown}
\end{figure}

\begin{figure}[]
   \centering
    \includegraphics[width=\columnwidth]{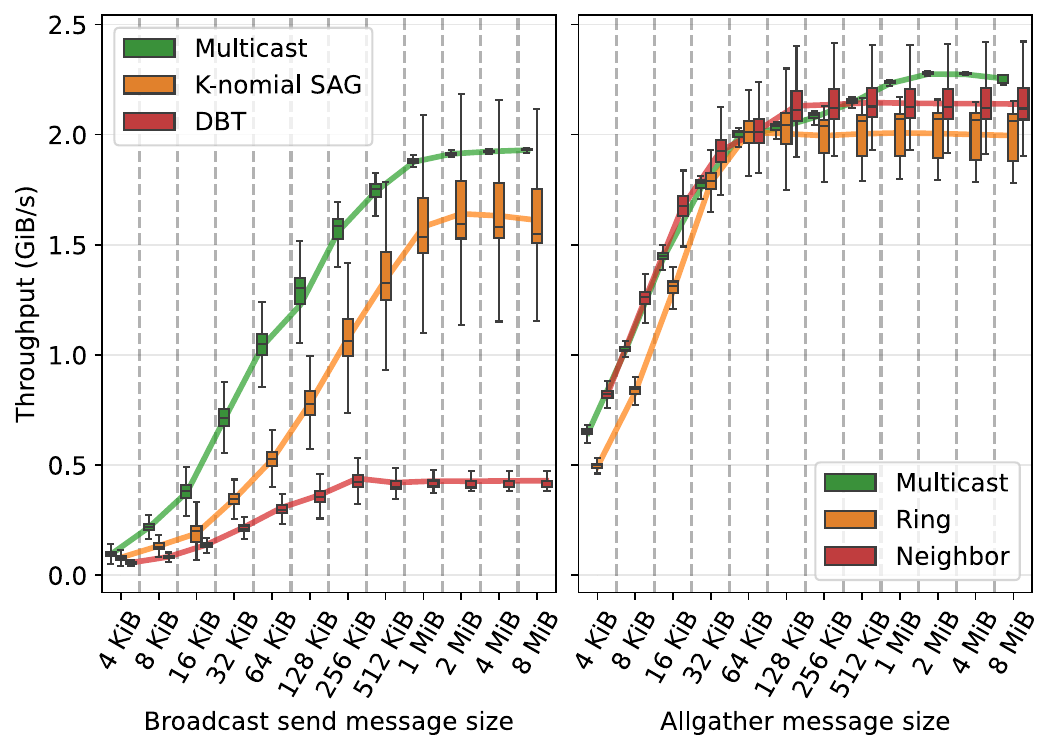}
    \caption{The per-process receive throughput at 188 nodes.}
    \label{fig:osu}
\end{figure}

\begin{figure}[]
   \centering
    \includegraphics[width=\columnwidth]{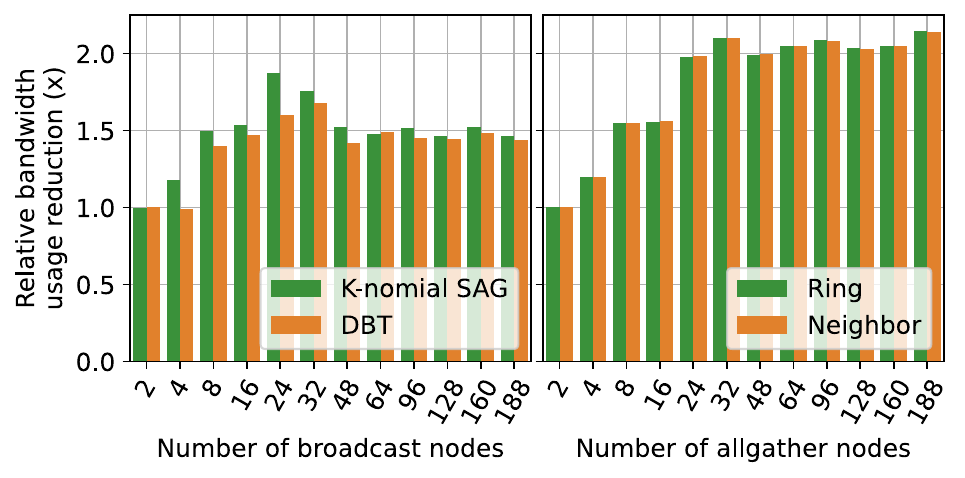}
    \caption{Multicast-based algorithms achieve up to $2\times$ traffic reduction across 18 SX6036 switches, when compared to P2P-based Broadcast and Allgather algorithms.}
    \label{fig:switch_port_counters}
\end{figure}

We focus on intra-node communication and investigate the collective stack performance with 1 process per node (PPN). We allocate 1 worker thread per send and receive datapaths. As baseline algorithms, we choose the bandwidth-optimized collective algorithms implemented in the UCC library using the P2P UCX backend~\cite{ucc, ucx}. We perform 100 warm-up iterations that are excluded from reported measurements. We report per-rank performance across 1000 iterations for throughput experiments with send message size below 4096 bytes and 100 iterations for larger messages. For Broadcast, we report measurements only on leaf ranks. We use one actively multicasting root within the Allgather schedule.

\paragraph{Protocol scalability} We designed the protocol for maximum communication/computation overlap. We investigate the synchronization overheads in our protocol (i.e., RNR barrier at the initialization stage and the final handshake) in Figure~\ref{fig:critical_path_breakdown}. As the message size and number of participants grow, the non-blocking multicast datapath starts to dominate. Starting from 16 nodes, 99\% of CPU time in the Allgather progress path is spent in the data movement performed by send/receive worker threads. This suggests that our approach will be the most efficient at large system scales, as the protocol time will be dominated by the multicast.

\paragraph{Throughput at scale} Figure~\ref{fig:osu} shows the performance of our algorithms at the full system scale of 188 nodes. Our Broadcast algorithm outperforms k-nomial and binary tree P2P schemes by up to $1.3\times$ and $4.75\times$, respectively. For 128 --- 256 KiB Allgather, typical for FSDP training~\cite{pytorch, nccl, rccl}, the multicast approach achieves the same throughput as the ring algorithm. Such alignment is expected as the throughput of both algorithms is bounded by the receive buffer size~\cite{thakur2003mpichcolls}.

\paragraph{Performance variability} As the message size grows, we also notice significantly smaller throughput variability in multicast-based collectives than in P2P algorithms. For large send buffers, the latency of RNR synchronization becomes negligible (see Figure~\ref{fig:critical_path_breakdown}), and the communication is dominated by the single-root multicasting.

\paragraph{Traffic savings} The goal of our algorithms is to minimize the data movement across the network. By doing so, we can free the injection bandwidth for other in-flight operations. We assess the traffic reduction in Figure~\ref{fig:switch_port_counters}. We collect performance counters across all switch ports of our Fat-Tree testbed when running the Broadcast and Allgather collectives. We run 10 iterations with 64 KiB send message size for each measurement to ensure that counters do not overflow across iterations. With multicast-based Allgather, we achieve $1.5\times$ to $2\times$ data movement savings when compared to the P2P algorithms. This experiment confirms our theoretical insights from Section~\ref{sec:background} that multicast-based algorithms can minimize the data movement across the network.

\subsection{SmartNIC offloading}\label{sec:dpa_perf}

We study the performance scalability of our multi-threaded Allgather receive progress engine offloaded to the DPA.

\paragraph{Experimental setup} In Figure~\ref{fig:critical_path_breakdown} we assessed the protocol efficiency across the testbed scale. We notice that with small scale/message sizes, latency in the RNR and Final synchronization dominates the CPU cycles, becoming negligible only at larger scale. In the DPA experiments, we assume the scenario where our protocol is deployed at a large scale, and Allgather progress engine time is predominantly spent in the datagram reception, which we offload to the DPA. 

We reproduce a scenario where the receive datapath (DPA server) acts as a leaf node in the schedule of our Allgather algorithm. The leaf is fully saturated with datagrams coming from the x86 client machine that simulates the broadcasting roots. To imitate the traffic distribution across multicast trees, we create multiple QP connections between the client and the server. Each receive DPA worker processes the traffic from one connection (multicast tree). We co-locate receive threads on DPA in a compact way so that first, we occupy 16 hardware threads of core 1, then core 2, etc. By doing so we exercise the ability of DPA to sustain the load when all worker threads share the same core resource. UD-based datapath uses BlueField 3 DRAM banks for staging area allocation.

\paragraph{Baseline datapath} Our baseline setup is intended to simulate conventional CPU-based datagram processing. Similarly to UC-based DPA datapath, the baseline receiver performs a logical zero-copy re-assembly of received packets. With such a baseline, we can estimate a practical lower bound on the performance of single-threaded CPU-based buffer processing with per-datagram granularity.

\begin{figure}[]
   \centering
    \includegraphics[width=\columnwidth]{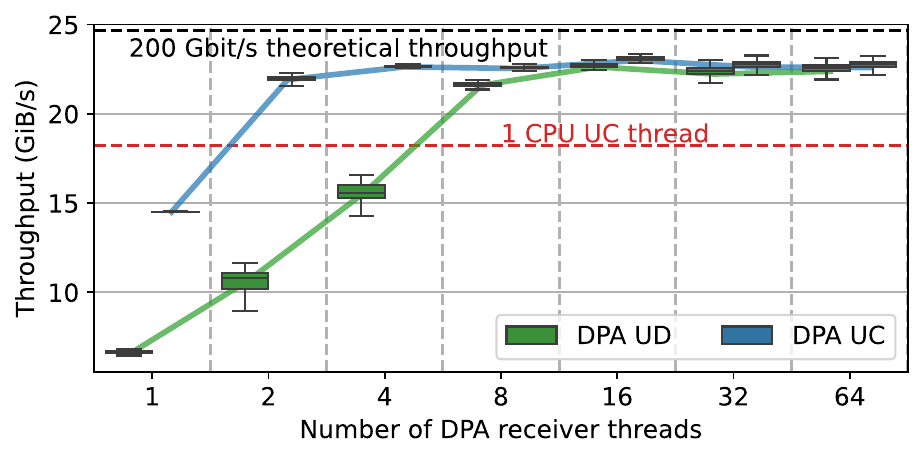}
    \caption{The throughput scaling with 8 MiB receive buffer size and 4 KiB chunk size as the number of DPA threads increases.}
    \label{fig:dpa_thread_num_scaling}
\end{figure}

\begin{table}[]
\resizebox{\columnwidth}{!}{%
\begin{tabular}{@{}ccccc@{}}
\toprule
\textbf{Receive datapath} & \textbf{Throughput, GiB/s} & \textbf{Instructions/CQE} & \textbf{Cycles/CQE} & \textbf{IPC} \\ \cmidrule{1-5}
UC         & 11.9 & 66  & 598  & 0.11 \\ \addlinespace[0.5em]
UD         & 5.2  & 113 & 1084 & 0.1  \\ \bottomrule
\end{tabular}%
}
\caption{Average DPA single-thread performance metrics with 8 MiB receive buffer and 4 KiB chunk size.}
\label{tab:dpa_single_thread_perf}
\end{table}

\begin{figure}[]
   \centering
    \includegraphics[width=\columnwidth]{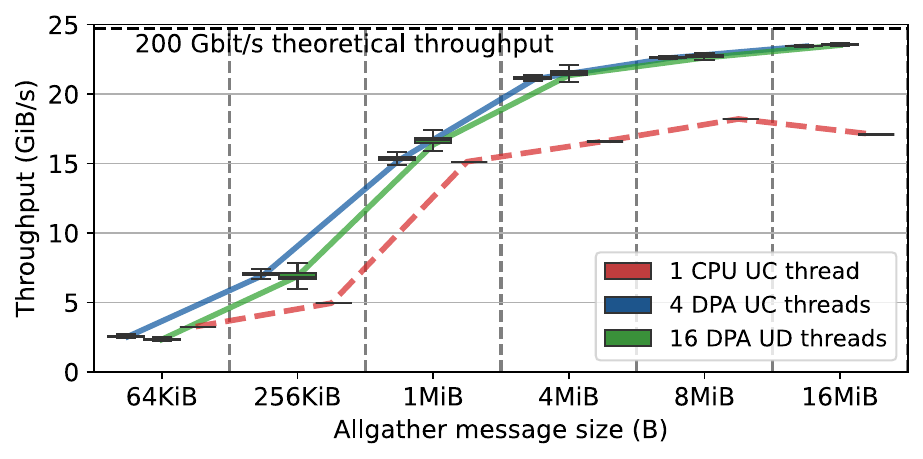}
    \caption{The DPA throughput scaling with 4 KiB chunk size.}
    \label{fig:dpa_message_size_scaling}
\end{figure}

\begin{figure}[]
    \centering
    \includegraphics[width=\columnwidth]{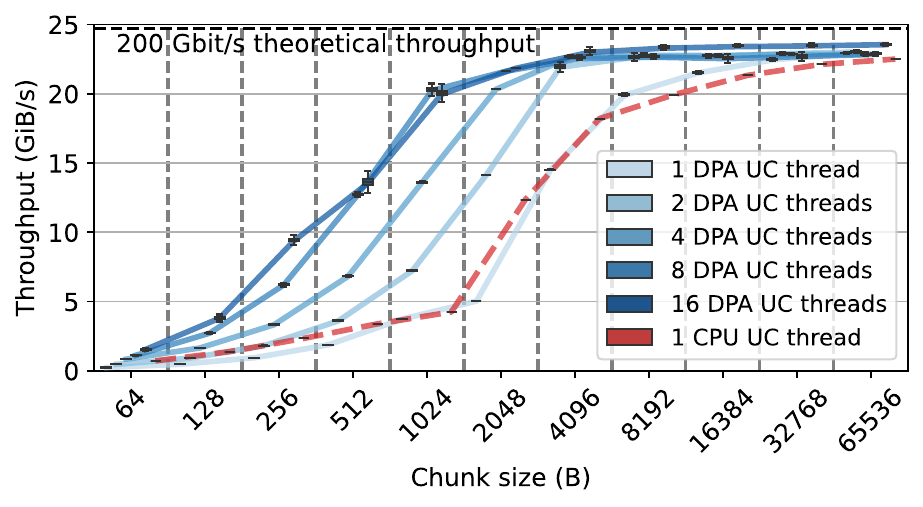}
    \caption{Throughput of UC transport with 8 MiB send buffer and multi-packet chunk sizes.}
    \label{fig:dpa_chunk_size_uc}
\end{figure}

\paragraph{Single thread performance} In Table~\ref{tab:dpa_single_thread_perf}, we investigate the single-threaded receive datapath performance metrics. With $1/256$ of DPA capacity, the datapaths achieve $1/2$ (UC) and $1/5$ (UD) of peak theoretical throughput (200 Gbit/s). The equally low IPC metric for both datapaths suggests that the processing is bottlenecked by high-latency load/store operations (RDMA operations re-posting and bitmap updates).

\paragraph{Receive side scaling} We scale the number of DPA threads to hide the latency of single-thread processing. In Figures~\ref{fig:dpa_thread_num_scaling} and~\ref{fig:dpa_message_size_scaling}, each thread is mapped to an independent connection and processes different blocks of the receive buffer. UC datapath reaches the full throughput with 4 threads, whereas a UD-based receiver that has $\approx2\times$ higher critical-path latency needs from 8 to 16 threads. Notably, with the number of hardware threads that fit into \textit{only 1 DPA core} ($1/16$ of total DPA capacity), both datapaths reach the practical link throughput and outperform single CPU core by $25\%$.

\paragraph{Improvements of UC-based multicast}
UC transport supports \textit{arbitrary length} RDMA Write messages. With UC multicast we can further minimize the software involvement on the receive side, the multi-packet messages will arrive with the less frequency when compared to per-datagram processing. In Figure~\ref{fig:dpa_chunk_size_uc}, we investigate the impact of the chunk size on the processing of an 8 MiB buffer. With the larger chunk size, DPA can sustain a line rate with fewer threads. Multi-packet UC multicast combined with our scheduling in Allgather protocol will result in a low software-overhead solution.

\section{Applicability}

\begin{figure}[]
    \centering
    \includegraphics[width=\columnwidth]{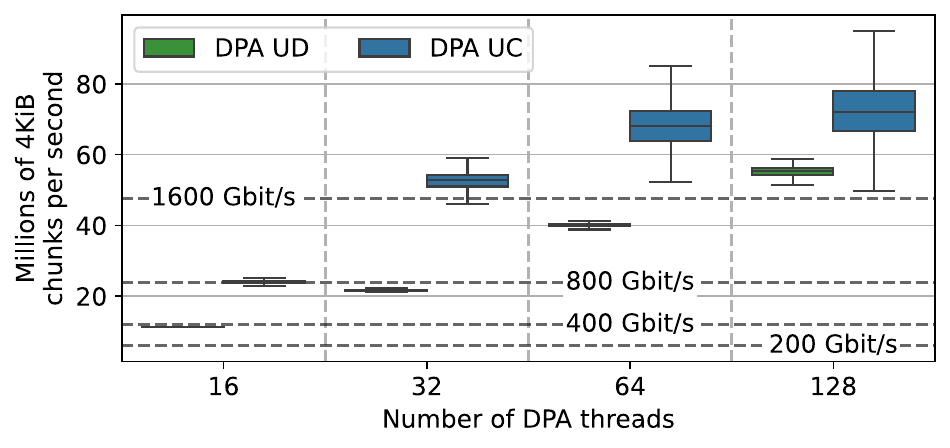}
    \caption{Sustained chunk processing rate with the DPA-offloaded collective receive datapath. The number of hardware threads is scaled up to half of the total DPA compute capacity.}
    \label{fig:1600_scaling}
\end{figure}

\paragraph{Scaling to next-generation Tbit/s links} Ethernet link speeds are projected to reach Tbit/s speeds by 2025~\cite{ieee_dambrosia, hoefler2023data}. In Section~\ref{sec:evaluation}, we show that a single DPA core running UD-based Allgather can sustain 200 Gbit/s bandwidth while utilizing $1/16$'th of the full DPA compute capacity. We are now studying the scalability of Allgather's receive datapath for future link bandwidths. Assuming $N\times$ higher link bandwidth and 100\% link utilization, MTU-sized packets will arrive at the receive side at $\approx N\times$ higher rate. To investigate the datapath potential to sustain higher rates, we decrease the chunk size in the fragmentation protocol to 64 Bytes to match the chunk arrival rate of a 1600 Gbit/s network with 4096 KiB MTU-sized packets. Figure~\ref{fig:1600_scaling} shows the sustained chunk processing rate for UD and UC using different numbers of DPA threads. With 128 threads (i.e., half of DPA cores), our approach sustains 1600 Gbit/s already on current-generation DPAs.

\paragraph{Protocol deployment scope}
The $2\times$ traffic reduction achieved with our multicast-based Allgather algorithm makes it a promising solution for next-generation systems leveraging FSDP training. As discussed in Section~\ref{sec:background}, collective operations competing with Allgather for network injection bandwidth (e.g., Reduce-Scatter) can achieve lower completion times. Our constant-time Broadcast algorithm, used in conjunction with SmartNIC-based offloading and fabric Quality-of-Service management mechanisms (e.g., Virtual Lanes \cite{alfaro2004qos}), would perfectly fit the applications imposing stringent requirements on the completion time of network communication, such as distributed file system replication \cite{di2022building,kim2021linefs}.

\paragraph{DPA offloading for other workloads}
Our DPA-based solution is entirely software-defined. The Allgather and Broadcast kernels we developed can be expressed with any Verbs-like API that exposes NIC RDMA engine capabilities to the user. In this context, our UD-based datapath offloading methodology that, by design, supports out-of-order packet arrival can be applied to RoCEv2/UDP traffic. We believe that other software-defined protocols featuring traffic parallelism and requiring reliable data transmission, such as QUIC \cite{langley2017quic}, and storage I/O \cite{min2021gimbal, pismenny2021autonomous}, can ideally fit DPA offloading.

\paragraph{\hl{Economics of SmartNIC offloading}}
The single-threaded CPU-based baseline in Figures~\ref{fig:cpu_bottleneck},~\ref{fig:dpa_thread_num_scaling} shows that one CPU core can sustain datagrams receiving at only $\approx 1/2$ to $2/3$ of a 200 Gbit/s link. Thus, the progress engine core count footprint will at least double with twice higher link bandwidth. For example, to saturate four 1.6 Tbit/s links with 4 KiB datagrams in both directions, we will need at least 64 CPU cores (1 core per 100 Gbit/s). In Figure~\ref{fig:1600_scaling}, we demonstrate that the DPA generation we utilize in this work already has enough compute power to drive such a link. Let’s consider the NVIDIA SuperPOD node based on the $2\times$ 54-core Xeon 8570 CPU and $4\times$ ConnectX-7 400 Gbit/s NICs supporting DPA offloading. When compared to the CPUs, the NICs total cost and energy consumption are $\approx 2.5\times$ lower and $\approx 7\times$ lower, correspondingly.

\hl{In this light}, DPA offloading and on-path SmartNIC offloading in general (as opposed to the off-path Linux-based SoC offloading deployed for cloud infrastructure) can be a cost-efficient alternative to classical CPU-driven systems in:
\begin{enumerate}[leftmargin=*]
\item HPC machines specialized for energy-efficient FSDP training. Cheap CPUs with low core counts drive training \textit{control path}, while compute and traffic processing \textit{kernels} are offloaded to energy-efficient GPUs and on-path SmartNICs.
\item General-purpose cloud HPC deployments where the expensive host resources (CPU cores, host-side interconnect) are shared across applications, e.g., storage and training.
\end{enumerate}

\section{Related Work}

To the best of our knowledge, our work is the first to propose a multicast-based Allgather algorithm and provide its open-source end-to-end implementation. Previous algorithmic approaches leveraging hardware multicast have primarily focused on Broadcast and reduction collectives.

The SHARP protocol relies on hardware multicast to perform in-network reduction~\cite{graham2016sharp}. Hoefler et al. ~\cite{hoefler2007bcast} introduced a constant-time Broadcast protocol for small messages. In Section~\ref{sec:broadcast}, we analyzed the reliability layer designs proposed in works ~\cite{chu2018gpudirectmvapich, chu2016mcastreliability}. The HCOLL library~\cite{hcoll} leverages multicast for the Broadcast collective, although it is closed-source. Related paper~\cite{dang2018lrum} suggests that the reliability protocol is built around a process tree with a sliding window between parent and child nodes. Our approach to reliability is orthogonal and does not require ACK'ing in the datapath. After RNR synchronization between participants, we rely on the multi-threaded receive path that provides enough processing capacity to sustain the line rate and minimize the probability of RNR drops. In case drops happen, we rely on selective fetching in the reliable ring. Receive side scaling was extensively studied by  Belay and Prekas~\cite{belay2014ix,prekas2017zygos,rss}.

The event-driven DPA programming model resembles ideas originating in active messaging~\cite{von1992active, ucx}, triggered operations in Portals 4~\cite{barrett2017portals, schneider2013protocols}, and the CORE-Direct feature~\cite{graham2010coredirect}. Existing implementations of these features are CPU-driven and do not support SmartNIC offloading.

Several works by the MVAPICH team focus on achieving non-blocking Alltoall collective progress~\cite{bayatpour2021bluesmpi, suresh2023novel}. These works focus on offloading a conventional single-threaded MPI progress engine to off-path general-purpose BlueField ARM cores. In contrast, our work takes a further step in understanding collective runtime co-design that exploits the DPA architecture, a part of the ConnectX complex tailored for highly parallel traffic processing.

\section{Conclusions}

In our work, we introduced novel bandwidth-optimal multicast-based Broadcast and Allgather algorithms that reduce data movement across the network by $1.5\times$ and $2\times$, respectively. We presented their open-source end-to-end implementation for standard RDMA interconnects. Furthermore, we demonstrated that the Datapath Accelerator is a suitable SmartNIC substrate for parallel collective progress engine offloading. We believe that our bandwidth-optimal collective algorithms, coupled with DPA-based offloading, present a promising solution for the next generation of \mbox{large-scale AI supercomputers with Tbit/s network links}.

\section*{Acknowledgments}
\hl{We thank NVIDIA for} hosting MK’s internship where the initial DPA prototype was developed. We also thank CSCS for providing computational resources for this project. This project received funding from ERC PSAP (grant 101002047) and a donation from Intel.

\bibliographystyle{ieeetr}
\bibliography{references}

\begin{thebibliography}{10}

\bibitem{zhao2023pytorch}
Y.~Zhao, A.~Gu, R.~Varma, L.~Luo, C.-C. Huang, M.~Xu, L.~Wright, H.~Shojanazeri, M.~Ott, S.~Shleifer, {\em et~al.}, ``Pytorch fsdp: experiences on scaling fully sharded data parallel,'' {\em arXiv preprint arXiv:2304.11277}, 2023.

\bibitem{rasley2020deepspeed}
J.~Rasley, S.~Rajbhandari, O.~Ruwase, and Y.~He, ``Deepspeed: System optimizations enable training deep learning models with over 100 billion parameters,'' in {\em Proceedings of the 26th ACM SIGKDD International Conference on Knowledge Discovery \& Data Mining}, pp.~3505--3506, 2020.

\bibitem{rajbhandari2020zero}
S.~Rajbhandari, J.~Rasley, O.~Ruwase, and Y.~He, ``Zero: Memory optimizations toward training trillion parameter models,'' in {\em SC20: International Conference for High Performance Computing, Networking, Storage and Analysis}, pp.~1--16, IEEE, 2020.

\bibitem{dubey2024llama}
A.~Dubey, A.~Jauhri, A.~Pandey, A.~Kadian, A.~Al-Dahle, A.~Letman, A.~Mathur, A.~Schelten, A.~Yang, A.~Fan, {\em et~al.}, ``The llama 3 herd of models,'' {\em arXiv preprint arXiv:2407.21783}, 2024.

\bibitem{walker1996mpi}
D.~W. Walker and J.~J. Dongarra, ``{MPI: a standard message passing interface},'' {\em Supercomputer}, vol.~12, pp.~56--68, 1996.

\bibitem{nccl}
{Nvidia}, ``{NVIDIA Collective Communications Library (NCCL)}.'' \url{https://developer.nvidia.com/nccl}.

\bibitem{oneccl}
{Intel}, ``{Intel oneAPI Collective Communications Library}.'' \url{https://www.intel.com/content/www/us/en/developer/tools/oneapi/oneccl.html}.

\bibitem{ucc}
{Unified Communication Framework (UCF) consortium}, ``{Unified Collective Communication (UCC)}.'' \url{https://pytorch.org}.

\bibitem{rccl}
{AMD}, ``{ROCm Collective Communication Library (RCCL)}.'' \url{https://rocm.docs.amd.com/projects/rccl/en/latest/}.

\bibitem{zhou2023accelerating}
Q.~Zhou, Q.~Anthony, L.~Xu, A.~Shafi, M.~Abduljabbar, H.~Subramoni, and D.~K.~D. Panda, ``Accelerating distributed deep learning training with compression assisted allgather and reduce-scatter communication,'' in {\em 2023 IEEE International Parallel and Distributed Processing Symposium (IPDPS)}, pp.~134--144, IEEE, 2023.

\bibitem{thakur2003mpichcolls}
R.~Thakur and W.~D. Gropp, ``{Improving the performance of collective operations in MPICH},'' in {\em European Parallel Virtual Machine/Message Passing Interface Users’ Group Meeting}, pp.~257--267, Springer, 2003.

\bibitem{chan2007collective}
E.~Chan, M.~Heimlich, A.~Purkayastha, and R.~Van De~Geijn, ``Collective communication: theory, practice, and experience,'' {\em Concurrency and Computation: Practice and Experience}, vol.~19, no.~13, pp.~1749--1783, 2007.

\bibitem{alexandrov1995loggp}
A.~Alexandrov, M.~F. Ionescu, K.~E. Schauser, and C.~Scheiman, ``{LogGP: Incorporating long messages into the LogP model—one step closer towards a realistic model for parallel computation},'' in {\em Proceedings of the seventh annual ACM symposium on Parallel algorithms and architectures}, pp.~95--105, 1995.

\bibitem{belay2014ix}
A.~Belay, G.~Prekas, A.~Klimovic, S.~Grossman, C.~Kozyrakis, and E.~Bugnion, ``{IX: a protected dataplane operating system for high throughput and low latency},'' in {\em 11th USENIX Symposium on Operating Systems Design and Implementation (OSDI 14)}, pp.~49--65, 2014.

\bibitem{danalis2005transformations}
A.~Danalis, K.-Y. Kim, L.~Pollock, and M.~Swany, ``Transformations to parallel codes for communication-computation overlap,'' in {\em SC'05: Proceedings of the 2005 ACM/IEEE conference on Supercomputing}, pp.~58--58, IEEE, 2005.

\bibitem{ibspec}
{InfiniBand Trade Association}, ``{InfiniBand Specification}.'' \url{https://www.infinibandta.org}.

\bibitem{hoefler2017spin}
T.~Hoefler, S.~Di~Girolamo, K.~Taranov, R.~E. Grant, and R.~Brightwell, ``{sPIN: High-performance streaming Processing in the Network},'' in {\em Proceedings of the International Conference for High Performance Computing, Networking, Storage and Analysis}, pp.~1--16, 2017.

\bibitem{chen2024demystifying-bf3dpa}
X.~Chen, J.~Zhang, T.~Fu, Y.~Shen, S.~Ma, K.~Qian, L.~Zhu, C.~Shi, M.~Liu, and Z.~Wang, ``{Demystifying Datapath Accelerator Enhanced Off-path SmartNIC},'' {\em arXiv preprint arXiv:2402.03041}, 2024.

\bibitem{bf3}
{NVIDIA}, ``{Bluefield-3 DPU}.'' \url{https://resources.nvidia.com/en-us-accelerated-networking-resource-library/datasheet-nvidia-bluefield}.

\bibitem{li2020pytorch}
S.~Li, Y.~Zhao, R.~Varma, O.~Salpekar, P.~Noordhuis, T.~Li, A.~Paszke, J.~Smith, B.~Vaughan, P.~Damania, {\em et~al.}, ``Pytorch distributed: Experiences on accelerating data parallel training,'' {\em arXiv preprint arXiv:2006.15704}, 2020.

\bibitem{ben2019demystifying}
T.~Ben-Nun and T.~Hoefler, ``Demystifying parallel and distributed deep learning: An in-depth concurrency analysis,'' {\em ACM Computing Surveys (CSUR)}, vol.~52, no.~4, pp.~1--43, 2019.

\bibitem{kurth2017deep}
T.~Kurth, J.~Zhang, N.~Satish, E.~Racah, I.~Mitliagkas, M.~M.~A. Patwary, T.~Malas, N.~Sundaram, W.~Bhimji, M.~Smorkalov, {\em et~al.}, ``Deep learning at 15pf: supervised and semi-supervised classification for scientific data,'' in {\em Proceedings of the International Conference for High Performance Computing, Networking, Storage and Analysis}, pp.~1--11, 2017.

\bibitem{al2008scalable}
M.~Al-Fares, A.~Loukissas, and A.~Vahdat, ``A scalable, commodity data center network architecture,'' {\em ACM SIGCOMM computer communication review}, vol.~38, no.~4, pp.~63--74, 2008.

\bibitem{gangidi2024rdma}
A.~Gangidi, R.~Miao, S.~Zheng, S.~J. Bondu, G.~Goes, H.~Morsy, R.~Puri, M.~Riftadi, A.~J. Shetty, J.~Yang, {\em et~al.}, ``{RDMA over Ethernet for Distributed Training at Meta Scale},'' in {\em Proceedings of the ACM SIGCOMM 2024 Conference}, pp.~57--70, 2024.

\bibitem{cscs_alps}
CSCS, ``{CSCS, Hewlett Packard Enterprise and NVIDIA Announce World’s Most Powerful AI-Capable Supercomputer},'' {\em Swiss National Supercomputing Center}.

\bibitem{meta_ai_center_wagner}
K.~Wagner, ``{Meta Is Building New \$800 Million AI-Focused Data Center in Indiana},'' {\em Bloomberg}.

\bibitem{de2020depth}
D.~De~Sensi, S.~Di~Girolamo, K.~H. McMahon, D.~Roweth, and T.~Hoefler, ``An in-depth analysis of the slingshot interconnect,'' in {\em SC20: International Conference for High Performance Computing, Networking, Storage and Analysis}, pp.~1--14, IEEE, 2020.

\bibitem{li2019evaluating}
A.~Li, S.~L. Song, J.~Chen, J.~Li, X.~Liu, N.~R. Tallent, and K.~J. Barker, ``Evaluating modern gpu interconnect: Pcie, nvlink, nv-sli, nvswitch and gpudirect,'' {\em IEEE Transactions on Parallel and Distributed Systems}, vol.~31, no.~1, pp.~94--110, 2019.

\bibitem{hoefler2023data}
T.~Hoefler, D.~Roweth, K.~Underwood, R.~Alverson, M.~Griswold, V.~Tabatabaee, M.~Kalkunte, S.~Anubolu, S.~Shen, M.~McLaren, {\em et~al.}, ``Data center ethernet and remote direct memory access: Issues at hyperscale,'' {\em Computer}, vol.~56, no.~7, pp.~67--77, 2023.

\bibitem{dang2018lrum}
H.-V. Dang, B.~Smith, R.~Graham, and G.~Shainer, ``{LRUM: Local reliability protocol for unreliable hardware multicast},'' in {\em Proceedings of the International Conference on High Performance Computing in Asia-Pacific Region}, pp.~210--221, 2018.

\bibitem{di2022building}
S.~Di~Girolamo, D.~De~Sensi, K.~Taranov, M.~Malesevic, M.~Besta, T.~Schneider, S.~Kistler, and T.~Hoefler, ``Building blocks for network-accelerated distributed file systems,'' in {\em SC22: International Conference for High Performance Computing, Networking, Storage and Analysis}, pp.~1--14, IEEE, 2022.

\bibitem{kim2021linefs}
J.~Kim, I.~Jang, W.~Reda, J.~Im, M.~Canini, D.~Kosti{\'c}, Y.~Kwon, S.~Peter, and E.~Witchel, ``Linefs: Efficient smartnic offload of a distributed file system with pipeline parallelism,'' in {\em Proceedings of the ACM SIGOPS 28th Symposium on Operating Systems Principles}, pp.~756--771, 2021.

\bibitem{koop2007zero}
M.~J. Koop, S.~Sur, and D.~K. Panda, ``Zero-copy protocol for mpi using infiniband unreliable datagram,'' in {\em 2007 IEEE International Conference on Cluster Computing}, pp.~179--186, IEEE, 2007.

\bibitem{chu2016mcastreliability}
C.-H. Chu, K.~Hamidouche, H.~Subramoni, A.~Venkatesh, B.~Elton, and D.~K. Panda, ``Efficient reliability support for hardware multicast-based broadcast in gpu-enabled streaming applications,'' in {\em 2016 First International Workshop on Communication Optimizations in HPC (COMHPC)}, pp.~29--38, IEEE, 2016.

\bibitem{docadma}
{NVIDIA}, ``{DOCA DMA}.'' \url{https://docs.nvidia.com/doca/sdk/doca+dma/index.html}.

\bibitem{agarwal2022understanding}
S.~Agarwal, R.~Agarwal, B.~Montazeri, M.~Moshref, K.~Elmeleegy, L.~Rizzo, M.~A. de~Kruijf, G.~Kumar, S.~Ratnasamy, D.~Culler, {\em et~al.}, ``Understanding host interconnect congestion,'' in {\em Proceedings of the 21st ACM Workshop on Hot Topics in Networks}, pp.~198--204, 2022.

\bibitem{kalia2016rdmadesign}
A.~Kalia, M.~Kaminsky, and D.~G. Andersen, ``{Design guidelines for high performance RDMA systems},'' in {\em 2016 USENIX Annual Technical Conference (USENIX ATC 16)}, pp.~437--450, 2016.

\bibitem{zambre2019breaking}
R.~Zambre, M.~Grodowitz, A.~Chandramowlishwaran, and P.~Shamis, ``Breaking band: A breakdown of high-performance communication,'' in {\em Proceedings of the 48th International Conference on Parallel Processing}, pp.~1--10, 2019.

\bibitem{docaflexio}
{NVIDIA}, ``{DOCA DPA Subsystem}.'' \url{https://docs.nvidia.com/doca/sdk/dpa+subsystem/index.html}.

\bibitem{hoefler2015scientific}
T.~Hoefler and R.~Belli, ``Scientific benchmarking of parallel computing systems: twelve ways to tell the masses when reporting performance results,'' in {\em Proceedings of the international conference for high performance computing, networking, storage and analysis}, pp.~1--12, 2015.

\bibitem{ucx}
{The Linux Foundation}, ``{Unified Communication X}.'' \url{https://openucx.org}.

\bibitem{pytorch}
{The Linux Foundation}, ``{PyTorch}.'' \url{https://pytorch.org}.

\bibitem{ieee_dambrosia}
J.~D’Ambrosia, ``{IEEE P802.3df Defines Architecture Holistically to Achieve 800 Gb/s and 1.6 Tb/s Ethernet},'' {\em IEEE Standards Association}, 2022.

\bibitem{alfaro2004qos}
F.~J. Alfaro, J.~L. S{\'a}nchez, and J.~Duato, ``{QoS in InfiniBand subnetworks},'' {\em IEEE Transactions on Parallel and Distributed Systems}, vol.~15, no.~9, pp.~810--823, 2004.

\bibitem{langley2017quic}
A.~Langley, A.~Riddoch, A.~Wilk, A.~Vicente, C.~Krasic, D.~Zhang, F.~Yang, F.~Kouranov, I.~Swett, J.~Iyengar, {\em et~al.}, ``The quic transport protocol: Design and internet-scale deployment,'' in {\em Proceedings of the conference of the ACM special interest group on data communication}, pp.~183--196, 2017.

\bibitem{min2021gimbal}
J.~Min, M.~Liu, T.~Chugh, C.~Zhao, A.~Wei, I.~H. Doh, and A.~Krishnamurthy, ``Gimbal: enabling multi-tenant storage disaggregation on smartnic jbofs,'' in {\em Proceedings of the 2021 ACM SIGCOMM 2021 Conference}, pp.~106--122, 2021.

\bibitem{pismenny2021autonomous}
B.~Pismenny, H.~Eran, A.~Yehezkel, L.~Liss, A.~Morrison, and D.~Tsafrir, ``{Autonomous NIC offloads},'' in {\em Proceedings of the 26th ACM International Conference on Architectural Support for Programming Languages and Operating Systems}, pp.~18--35, 2021.

\bibitem{graham2016sharp}
R.~L. Graham, D.~Bureddy, P.~Lui, H.~Rosenstock, G.~Shainer, G.~Bloch, D.~Goldenerg, M.~Dubman, S.~Kotchubievsky, V.~Koushnir, {\em et~al.}, ``{Scalable hierarchical aggregation protocol (SHArP): A hardware architecture for efficient data reduction},'' in {\em 2016 First International Workshop on Communication Optimizations in HPC (COMHPC)}, pp.~1--10, IEEE, 2016.

\bibitem{hoefler2007bcast}
T.~Hoefler, C.~Siebert, and W.~Rehm, ``{A practically constant-time MPI Broadcast Algorithm for large-scale InfiniBand Clusters with Multicast},'' in {\em 2007 IEEE International Parallel and Distributed Processing Symposium}, pp.~1--8, IEEE, 2007.

\bibitem{chu2018gpudirectmvapich}
C.-H. Chu, X.~Lu, A.~A. Awan, H.~Subramoni, B.~Elton, and D.~K. Panda, ``{Exploiting hardware multicast and GPUDirect RDMA for efficient broadcast},'' {\em IEEE Transactions on Parallel and Distributed Systems}, vol.~30, no.~3, pp.~575--588, 2018.

\bibitem{hcoll}
{NVIDIA}, ``{HCOLL}.'' \url{https://docs.nvidia.com/networking/display/hpcxv216/hcoll}.

\bibitem{prekas2017zygos}
G.~Prekas, M.~Kogias, and E.~Bugnion, ``Zygos: Achieving low tail latency for microsecond-scale networked tasks,'' in {\em Proceedings of the 26th Symposium on Operating Systems Principles}, pp.~325--341, 2017.

\bibitem{rss}
{Microsoft}, ``{Introduction to Receive Side Scaling}.'' \url{https://learn.microsoft.com/en-us/windows-hardware/drivers/network/introduction-to-receive-side-scaling}.

\bibitem{von1992active}
T.~Von~Eicken, D.~E. Culler, S.~C. Goldstein, and K.~E. Schauser, ``Active messages: a mechanism for integrated communication and computation,'' {\em ACM SIGARCH Computer Architecture News}, vol.~20, no.~2, pp.~256--266, 1992.

\bibitem{barrett2017portals}
B.~Barrett, R.~B. Brightwell, R.~Grant, K.~Pedretti, K.~Wheeler, K.~D. Underwood, R.~Riesen, A.~B. Maccabe, T.~Hudson, and S.~Hemmert, ``{The Portals 4.1 network programming interface},'' tech. rep., Sandia National Lab.(SNL-NM), Albuquerque, NM (United States), 2017.

\bibitem{schneider2013protocols}
T.~Schneider, T.~Hoefler, R.~E. Grant, B.~W. Barrett, and R.~Brightwell, ``Protocols for fully offloaded collective operations on accelerated network adapters,'' in {\em 2013 42nd International Conference on Parallel Processing}, pp.~593--602, IEEE, 2013.

\bibitem{graham2010coredirect}
R.~L. Graham, S.~Poole, P.~Shamis, G.~Bloch, N.~Bloch, H.~Chapman, M.~Kagan, A.~Shahar, I.~Rabinovitz, and G.~Shainer, ``{Overlapping computation and communication: Barrier algorithms and ConnectX-2 CORE-Direct capabilities},'' in {\em 2010 IEEE International Symposium on Parallel \& Distributed Processing, Workshops and Phd Forum (IPDPSW)}, pp.~1--8, IEEE, 2010.

\bibitem{bayatpour2021bluesmpi}
M.~Bayatpour, N.~Sarkauskas, H.~Subramoni, J.~Maqbool~Hashmi, and D.~K. Panda, ``Bluesmpi: Efficient mpi non-blocking alltoall offloading designs on modern bluefield smart nics,'' in {\em International Conference on High Performance Computing}, pp.~18--37, Springer, 2021.

\bibitem{suresh2023novel}
K.~K. Suresh, B.~Michalowicz, B.~Ramesh, N.~Contini, J.~Yao, S.~Xu, A.~Shafi, H.~Subramoni, and D.~Panda, ``A novel framework for efficient offloading of communication operations to bluefield smartnics,'' in {\em 2023 IEEE International Parallel and Distributed Processing Symposium (IPDPS)}, pp.~123--133, IEEE, 2023.

\end{thebibliography}

\appendices

\newpage

\section{Distributed Broadcast sequencer}\label{sec:scheduler_appendix}

We formally define the multicast scheduling algorithm as follows. Let $P$ be the total number of processes participating in Allgather. Let $M$ be the size of a broadcasting group $G$, so that $P \mod M = 0$. We enumerate all processes from $0$ to $P-1$. We introduce the chain length as $R=P/M$, which equals the number of steps in the schedule. Assuming that the schedule steps are enumerated from $0$, at step $i$, the active group $G^{i}$ contains the processes: $$G^{i}=\{P_{i}, P_{R + i}, P_{2R + i}, ... ,  P_{(M - 1)R + i}\}$$

\section{Time reduction with multicast-based Allgather and INC Reduce-Scatter}\label{sec:ag_cost_appendix}
\hl{We denote a tuple of concurrent} Allgather ($AG$) and Reduce-Scatter ($RS$) collectives across $P$ processes as $\{AG, RS\}_{P}$. Both operations share the same network resources with full-duplex NICs and non-blocking fabric. We denote bandwidth towards a single NIC direction as $B_{nic}$. 

$N$ represents the send buffer size of a single $AG$ rank, which is equal to the receive buffer size of a single $RS$ rank. Concurrently, each $AG$ needs to receive $N(P-1)$ bytes from the network, which is equal to the $RS$ send buffer.

We assume that the latency of the first and last packets in the collective schedule is negligible (i.e., $N$ is sufficiently large). Thus, the time $T$ of the collective operation is dominated by the transmission time of messages.

 We compare two configurations for $\{AG, RS\}$:
\begin{enumerate}[leftmargin=*]
    \item $\{AG_{ring}, RS_{ring}\}$: both collectives are implemented using the ring algorithm~\cite{thakur2003mpichcolls,gangidi2024rdma}.
    \item $\{AG_{mc}, RS_{inc}\}$: $AG$ is implemented using the multicast-based algorithm described in Sections~\ref{sec:broadcast},~\ref{sec:allgather_offloading}, and $RS$ is implemented with the in-network compute algorithm (e.g., using SHARP v3~\cite{graham2016sharp}).
\end{enumerate}

In the $\{AG_{ring}, RS_{ring}\}$ configuration, send/receive $B_{nic}$ is shared equally:
    \begin{multline}
    B_{send}^{AG_{ring}} = B_{recv}^{AG_{ring}} = B_{recv}^{RS_{ring}} = B_{send}^{RS_{ring}} = \frac{1}{2}B_{nic}.
    \end{multline}
For the $\{AG_{mc}, RS_{inc}\}$ setup, the $B_{nic}$ sharing is described as follows:
    \begin{equation}
    \begin{cases}
      B_{send}^{AG_{mc}} = B_{recv}^{RS_{inc}} = \frac{N}{N*(P-1)+N}B_{nic} = \frac{1}{P}B_{nic} \\
      B_{recv}^{AG_{mc}} = B_{send}^{RS_{inc}} = (1 - \frac{1}{P})B_{nic}
    \end{cases}\,.
\end{equation}

The speedup $S$ of $\{AG_{mc}, RS_{inc}\}_{P}$ over $\{AG_{ring},RS_{ring}\}_{P}$ can be calculated as:
\begin{multline}
S = \frac{T^{\{AG_{ring},RS_{ring}\}_{P}}}{T^{\{AG_{mc}, RS_{inc}\}_{P}}} = \frac{\frac{N(P-1)}{\frac{1}{2}B_{nic}}}{\frac{N(P-1)}{(1-\frac{1}{P})B_{nic}}} = 2 - \frac{2}{P}.
\end{multline}

\clearpage

\section{DPA-offloaded receive data path}\label{sec:dpa_appendix}
\begin{lstlisting}[float=*b, language=C, label={lst:dpa_uc_code}, caption=The simplified DPA C kernel for the UC-based receive datapath worker. The code leverages DOCA FlexIO API.]
static inline void dpa_tput_server_eh(struct dpa_export_data *export_data)
{
    /* Get DPA thread context */
    struct dpa_tput_ctx *serv_ctx = (struct dpa_tput_ctx *)export_data->app_ctx;
    uint64_t worker_id            = __atomic_fetch_add(&serv_ctx->worker_id, 1, __ATOMIC_RELAXED);
    flexio_uintptr_t qp_dbr_daddr = export_data->qp_transfer[worker_id].qp_dbr_daddr;
    uint64_t to_process           = serv_ctx->to_process;
    uint64_t *recvbuf_bitmap      = serv_ctx->recvbuf_bitmaps[worker_id];
    uint32_t last_recvd_chunk_id  = 0 - 1; // wrap it up to max uint64_t val
    uint32_t to_fetch             = 0;
    uint32_t recvd_chunk_id;
    uint64_t finish_place;
    uint32_t sq_pi;
    uint32_t cqes_consumed;
    uint32_t swqe_idx;
    flexio_uintptr_t host_counter_daddr;
    union flexio_dev_sqe_seg *swqe;
    struct flexio_dev_thread_ctx *dtctx;
    dpa_cq_ctx_t cq_ctx;
    flexio_dev_get_thread_ctx(&dtctx);
    /* Prepare CQ metadata for polling */
    cq_ctx_init(&cq_ctx,
                export_data->cq_transfer[worker_id].cq_num,
                export_data->cq_transfer[worker_id].log_cq_depth,
                export_data->cq_transfer[worker_id].ci_idx,
                export_data->cq_transfer[worker_id].cq_ring_daddr,
                export_data->cq_transfer[worker_id].cq_dbr_daddr,
                export_data->cq_transfer[worker_id].hw_owner_bit);
    /* Datapath loop */
    while (to_process) {
        if (flexio_dev_cqe_get_owner(cq_ctx.cqe) != cq_ctx.cq_hw_owner_bit) {
            if (flexio_dev_cqe_get_opcode(cq_ctx.cqe) != DPA_CQE_RESPONDER_WRITE_W_IMM) {
                return;
            }
            recvd_chunk_id = cqe_get_imm_data(cq_ctx.cqe);
            step_cq(&cq_ctx);
            rq_db_ring(qp_dbr_daddr, 1);
            bitmap_set_bit(recvbuf_bitmap, recvd_chunk_id);
            if (recvd_chunk_id - 1 != last_recvd_chunk_id)
                to_fetch++;
            last_recvd_chunk_id = recvd_chunk_id;
            to_process--;
        }
    }
    /* Update cached queue state */
    sq_pi = be32_to_cpu(*((uint32_t *)qp_dbr_daddr + 1));
    swqe_idx = get_wqe_idx(export_data->qp_transfer[worker_id].log_qp_sq_depth, sq_pi);
    swqe = (union flexio_dev_sqe_seg *)(export_data->qp_transfer[worker_id].qp_sq_daddr) + swqe_idx * 4;
    export_data->cq_transfer[worker_id].ci_idx += serv_ctx->to_process + 1;
    export_data->cq_transfer[worker_id].hw_owner_bit = cq_ctx.cq_hw_owner_bit;
    /* Notify host, e.g., the receive buffer can be released */
    finish_place = __atomic_fetch_add(&serv_ctx->finisher_id, 1, __ATOMIC_RELAXED);
    if (finish_place == (serv_ctx->n_workers - 1)) {
        flexio_dev_window_config(dtctx, export_data->window_id, serv_ctx->counter_lkey);
        flexio_dev_window_ptr_acquire(dtctx, serv_ctx->counter_ptr, &host_counter_daddr);
        *(uint64_t *)host_counter_daddr = DPA_WORK_COMPLETED_MAGICNUM + serv_ctx->iter;
        __dpa_thread_window_writeback();
    }
    /* Re-schedule DPA thread */
    flexio_dev_cq_arm(dtctx, cq_ctx.cq_idx, cq_ctx.cq_number);
    flexio_dev_thread_reschedule();
}
\end{lstlisting}

\end{document}